# Market force, ecology, and evolution

J. Doyne Farmer[1]

Prediction Company

Markets have internal dynamics leading to excess volatility and other phenomena that are difficult to explain using rational expectations models. This paper studies these using a nonequilibrium price formation rule, developed in the context of trading with market orders. Because this is so much simpler than a standard inter-temporal equilibrium model, it is possible to study multi-period markets analytically. There price dynamics have second order oscillatory terms. Value investing does not necessarily cause prices to track values. Trend following causes short term trends in prices, but also causes longer-term oscillations. When value investing and trend following are combined, even though there is little linear structure, there can be boom-bust cycles, excess and temporally correlated volatility, and fat tails in price fluctuations. The long term evolution of markets can be studied in terms of flows of money. Profits can be decomposed in terms of aggregate pairwise correlations. Under reinvestment of profits this leads to a capital allocation model that is equivalent to a standard model in population biology. An investigation of market efficiency shows that patterns created by trend followers are more resistant to efficiency than those created by value investors, and that profit maximizing behavior slows the progression to efficiency. Order of magnitude estimates suggest that the timescale for efficiency is years to decades[2].

---

1. Current address: Santa Fe Institute, 1399 Hyde Park Rd., Santa Fe NM 87501, jdf@santafe.edu, FAX: 505-982-0565.

2. Earlier versions of this paper were written in 1994 and 1998. The first public presentation was in June 1997 (see footnote 10). This version has been submitted to the *Journal of Economic Behavior and Organization.*

# Summary of notation

**TABLE 1. Summary of notation**

| symbol | meaning |
| --- | --- |
| $c_t^{(i)}$ | capital of $i^{th}$ agent or strategy at time $t$ (see Section 2.3) |
| $d_t$ | dividend paid by the asset at time $t$ |
| $D$ | excess demand (demand - supply) |
| $g_t^{(i)}$ | profit of agent $i$ at time $t$ (see equation (26)) |
| $G^{(ij)}$ | profit of agent $j$ due to agent $i$ (see equation (32)) |
| $(i, j)$ | superscripts denoting a particular agent or strategy |
| $K$ | total number of shares in the market |
| $m_t$ | logarithm of mispricing, $m_t = p_t - \nu_t$ |
| $N$ | number of different agents or strategies |
| $p_t$ | logarithm of midpoint price at time $t$ |
| $P_t$ | midpoint price at time $t$ |
| $r_t$ | log return at time $t$, i.e. $r_t = p_t - p_{t-1}$ |
| $R^{(i)}$ | return to $i^{th}$ agent or strategy, i.e. $R^{(i)} = \langle g^{(i)} \rangle / c^{(i)}$ |
| $T$ | Threshold for entering a position in a state-dependent value strategy (see Section 3.1.3) |
| $\nu_t^{(i)}$ | log of perceived value of agent $i$ at time $t$ (see Section 3.1) |
| $V$ | The position of a simple value strategy (see Section 3.1.1) |
| $x_t^{(i)}$ | position of agent or strategy $i$ at time $t$ |
| $\alpha$ | ratio of capital to liquidity |
| $\gamma^{(i)}$ | inflow rate of capital from outside the market to agent $i$ |
| $\Delta$ | time difference $\Delta y_t = y_t - y_{t-1}$ |
| $\eta_t$ | random change to perceived logarithm of value at time $t$ (see Section 3.1) |
| $\kappa_t^j$ | price sensitivity $\kappa_t^j = (1/\lambda) \langle \partial \omega_{t+1} / (\partial p_{t-j}) \rangle$ (see equation (40)) |
| $\lambda$ | liquidity |
| $\mu^{(i)}$ | mean profit caused by dividends or correlation to external noise (see equation (32)) |
| $\omega_t^{(i)}$ | order placed by agent or strategy $i$ at time $t$ (can be positive or negative) |
| $\omega_t$ | net order at time $t$, $\omega_t = \sum \omega_t^{(i)}$ |
| $\pi$ | pattern in log-returns, corresponding to non-zero expectation |
| $\rho_x(\tau)$ | autocorrelation of $x_t$ for time lag $\tau$, i.e. the correlation of $x_t$ and $x_{t-\tau}$ |
| $\sigma_x$ | standard deviation of $x_t$ |
| $\tau$ | threshold for exiting a position; also used to denote time lags. |
| $\xi_t$ | random change in price at time $t$ due to external information |

# 1. Introduction

The concept of equilibrium is central in economics. While equilibrium models have clearly been very useful in economics, real markets exhibit properties that they have difficulty explaining. Prices display *excess volatility*, i.e., they change more than rational mea-sures of value would lead one to expect (Shiller 1997). Most large price changes occur in the absence of anything that can clearly be labeled as "news" (Cutler et al. 1989). In a full rational expectations equilibrium there should be no speculative trading (Geanakoplos and Sebenius 1983, Milgrom and Stokey 1982, Geanakoplos 1992), whereas trading in foreign exchange markets exceeds a trillion dollars a day, and is at least a factor of 50 greater than the daily world GNP. Furthermore, price movements have temporally correlated or *clustered volatility* (Mandelbrot 1963, 1997, Engle 1982) and fat tails[3]. These facts are difficult to reconcile with rational expectations equilibrium.

This paper develops a simple nonequilibrium theory of price formation that naturally explains the internal dynamics of prices and markets. The central goals are: (1) to provide a treatment of price formation that is motivated by the structure of markets, and is simpler than standard inter-temporal equilibrium models; (2) to use this to study the price dynamics of common trading strategies; (3) to understand how profits and losses drive market selection; and (4) to study the progression toward market efficiency in a framework that does not effectively assume it at the outset.

The paper is divided into three chapters, somewhat whimsically called "Force", "Ecology", and "Evolution". The first of these chapters gives a precise definition for the "market force" exerted by buying and selling, by deriving a simple price formation rule within a model involving a market maker and directional traders submitting market orders.

The simplicity of the price formation rule developed in the first part enables the study of prices in an infinite period setting (in contrast to the usual one or two-period settings of most equilibrium models). The chapter titled "Ecology" works several examples for commonly used strategies, such as value investing and trend following. Since the market maker is built into the price formation process it is possible to study the price dynamics of strategies one at a time, which gives insight into their dynamics when they are combined. While value investing strategies induce negative short term autocorrelations in prices, and trend following strategies induce positive short term autocorrelations, over longer timescales their effects are more complicated. In a non-equilibrium setting there are inevitably second-order terms in the price dynamics. Combining value investors and trend followers leads to excess volatility and price oscillations corresponding to boom-bust cycles.

The simple price formation rule derived in the first chapter is special in that the profits and losses of a given agent or strategy can be decomposed in terms of pairwise correlations with the positions of other agents or strategies. The chapter titled "Evolution" develops a simple theory of capital allocation. This leads to a set of differential equations that describe the flow of capital. These equations are equivalent to the generalized Lotka-Volterra equations, which are the standard model for population dynamics in ecology. Like biological species, financial strategies can have competitive, symbiotic, or predator-prey relationships. The tendency of a market to become more efficient can be understood in terms of an evolutionary progression toward a richer and more complex set of financial strategies.

---

3. The distribution of large price fluctuations decreases as a power law on timescales less than a month. For reviews see Lux (1996), Mantegna and Stanley (1999), or Farmer (1999).

The desire to build financial theories based on more realistic assumptions has led to several new strands of literature, including psychological approaches to risk-taking behavior[4], evolutionary game theory[5], and agent-based modeling of financial markets[6]. Although substantially different in methods and style, these emerging sub-fields are all attempts to go beyond economic theories based on assumptions of equilibrium and efficiency. In particular, psychological models of financial markets focus on the manner in which human psychology influences the economic decision making process as an explanation of apparent departures from rationality. Evolutionary game theory studies the evolution and steady-state equilibria of populations of competing strategies in highly idealized settings. Agent-based models are meant to capture complex learning behavior and dynamics in financial markets using more realistic markets, strategies, and information structures.

Agent based models are often so complicated that analytic results are difficult to obtain. At the other extreme, evolutionary game theory models are often so unrealistic that it is not clear they bear any relation to real economies. The framework developed here can either be viewed as a simple setting for agent based modeling, or a more realistic setting for an evolutionary game. On one hand the model is not dramatically more complicated than the iterated prisoner's dilemma (Axelrod 1984), yet on the other hand it attempts to describe markets at some level of realism. It should hopefully provide a simple theoretical framework in which to investigate the impact of psychological behavior on price dynamics.

The goal here is not to formulate the most realistic possible market model, but rather to formulate the simplest model that is also reasonable. The purpose is to make a canonical model around which other non-equilibrium models can be viewed as refinements. By keeping the model simple it becomes possible to investigate many questions analytically, such as whether prices reflect values, how markets evolve on longer timescales, and under what circumstances and on what timescale evolution causes them to become efficient.

---

4. For examples of work in "behavioral economics", see De Bondt and Thaler (1995), Kahneman and Tversky (1979), and Shiller (1998).

5. For examples of work in evolutionary game theory applied to economics, see Friedman (1991, 1998), Samuelson (1998), and Weibull (1996). See also the literature on the El Farol bar problem/minority game (Arthur 1994, Challet and Zhang, 1997).

6. A few examples of agent based modeling include Arthur et al. (1997), Bouchaud and Cont (1998), Chiarella (1992), Caldarelli et al. (1997), Chan et al. (1998), Coche (1998), Iori (1999), Joshi (1998), LeBaron (1999), Levy et al. (1996), Lux (1997, 1998), Lux and Marchesi (1999), Reick (1994), Takayasu et al. (1992, 1997) and Youssefmir et al. (1998). For reviews see LeBaron (1999b) and Tesfatsion (1999).

## 2. Force

### 2.1 The need for a non-equilibrium approach to price formation

There are serious theoretical problems with the assumption of economic equilibrium. As articulated by Phelps (1991):

> "*The agents of equilibrium models are not simply rational creatures; they have somehow come to possess fantastic knowledge. The equilibrium premise raises obvious problems of knowledge: why should it be supposed that all the agents have hit upon the true model, and how did they manage to estimate it and conform to it more and more closely? There has always been a strand of thought, running from Morgenstern in the 1930s to Frydman in the present, that holds that we cannot hope to understand the major events in the life of an economy, and perhaps also its everyday behavior, without entertaining hypotheses of disequilibrium.*

See also Frydman and Phelps (1983).

We will consider the words *disequilibrium, nonequilibrium, and out-of-equilibrium* as synonyms. A typical approach to disequilibrium price formation uses a price adjustment process of the form

$$\frac{dP}{dt} = f(D(P)),$$ (Eq 1)

where $P$ is the price at time $t$, $D$ is the excess demand (i.e. demand - supply), and $f$ is an increasing function. In classic treatments of Walrasian tatonnement, agents signal their excess demand to each other and adjust prices, but only make transactions when prices reach equilibrium (e.g. Samuelson 1941, 1947, Walker 1996). This does not reflect the way trades are done in most modern financial markets. While more sophisticated models have been developed that allow transactions out of equilibrium, the main thrust of such work has generally been to determine stability conditions that justify convergence to equilibrium (e.g. Fisher 1983). Nonequilibrium behavior is viewed as a complication in the route to equilibrium. Temporary equilibrium models (e.g. Grandmont 1988) have been more successful in producing concrete results, and in recent years the study of out-of-equilibrium behavior has become unpopular.

The theory of rational expectations is built upon the assumptions of perfect competition, rational expectations, market clearing, agent optimization, and full knowledge of prices in advance of transactions. All of these assumptions are questionable. It is possible to make incremental departures from this framework, for example, by allowing imperfect competition in the form of asymmetric information (Grossman 1989), or assuming bounded rationality (Simon 1956, Sargeant 1993). Empirical studies show half lives for market clearing as long as a week (e.g. Hansch et al. 1998). And as a previous market practitioner, I can provide an existence proof against agent optimization by assuring the reader that my firm's trading is rarely, if ever, optimal.

The assumption that prices are known in advance of transactions is also problematic. In modern financial markets changes in the demand of individual agents are expressed in terms of *orders*. The two most common types of orders are market orders and limit orders. A *market order* is a request to transact immediately at the best available price. The fill price for small market orders is often quoted, so that it is known in advance, but for large market orders the fill price is unknown. In contrast, a *limit order* is a request to transact only at a given price or better. Thus the fill price is known, but the time of the transaction is unknown -- indeed the transaction may not be completed at all. In both cases there is uncertainty in either the time or the price of the transaction. This means that, barring miracles, individual transactions occur out of equilibrium.

## 2.2 Price formation model

The goal of this section is to derive a *market impact function* $\phi$ *(*sometimes also called a *price impact function)* relating order flow and prices.

### 2.2.1 Model framework

We assume there are two types of financial agents, who trade an asset (measured in units of shares) that can be converted to *money* (which can be viewed as a risk free asset paying no interest). The first type of agents are *directional traders*. They buy or sell by placing market orders, which are always filled. In the typical case that the buy and sell orders of the directional traders do not match, the excess is taken up by the second type of agent, who is a *market maker*. The orders are filled by the market maker at a price that is shifted from the previous price, by an amount that depends on the net order of the directional traders. Buying drives the price up, and selling drives it down. The market impact function $\phi$ is the algorithm that the market maker uses to set prices. This defines a price formation rule relating the net order to the new price.

Let there be $N$ directional traders, labeled by the superscript $i$, holding $x_t^{(i)}$ shares at time $t$. Although this is not necessary, in this paper we assume synchronous trading at times $\dots, t-1, t, t+1, \dots$. Let the position of the $i^{th}$ directional trader be a function[7] $x_{t+1}^{(i)} = X^{(i)}(P_t, P_{t-1}, \dots, I_t^{(i)})$, where $I_t^{(i)}$ represents any additional external information. The function $X^{(i)}$ can be thought of as the *strategy* or *decision rule* of agent $i$. The order $\omega_t^{(i)}$ is determined from the position through the relation

$$\omega_t^{(i)} = x_t^{(i)} - x_{t-1}^{(i)}. \tag{Eq 2}$$

A single timestep in the trading process can be decomposed into two parts:

1. The directional traders observe the most recent prices and information at time $t$ and submit orders $\omega_{t+1}^{(i)}$.

2. The market maker fills all the orders at the new price $P_{t+1}$

---

To keep things simple, we will assume that the price $P_t$ is a positive real number[8], and that positions, orders, and strategies are anonymous[9]. This motivates the assumption that the market maker bases price formation only on the net order

$$\omega = \sum_{i=1}^{N} \omega^{(i)}.$$

The algorithm the market maker uses to compute the fill price for the net order $\omega$ is an increasing function of $\omega$

$$P_{t+1} = f(P_t, \omega). \tag{Eq 3}$$

Note that because orders are anonymous, with more than one trader the fill price is unknown to individual agents when orders are placed.

### 2.2.2 Derivation of market impact function

An approximation of the market impact function can be derived by assuming that $f$ is of the form

$$f(P_t, \omega) = P_t \phi(\omega), \tag{Eq 4}$$

where $\phi$ is an increasing function with $\phi(0) = 1$. Taking logarithms and expanding in a Taylor's series, providing the derivative $\phi'(0)$ exists, to leading order[10]

$$\log P_{t+1} - \log P_t \approx \frac{\omega}{\lambda}. \tag{Eq 5}$$

This functional form for $\phi$ will be called the *log-linear* market impact function. $\lambda$ is a scale factor that normalizes the order size, and will be called the *liquidity*. It is the order size that will cause the price to change by a factor of $e$, and can be measured in either units of shares or money.

---

8. In real markets prices are not unique. Market makers are willing to buy at a low price, called the *bid*, or sell at a higher price, called the *offer* or the *ask*. The difference between the bid and the offer is called the *spread*. Requiring that prices are unique is equivalent to setting the spread to zero. This simplification neglects one of the primary sources of profits for market makers. Nonetheless, we will see that the market maker still makes profits on average, and the model is quite reasonable.

9. In real markets agents often use multiple brokers to disguise the identity of their orders, to keep trading anonymous. This can be inconvenient, and not all agents use all possible brokers, so it is often possible to make inferences about identity. Thus some traders may have a bigger influence on price formation than others (e.g. O'Hara (1995)).

10. This derivation, along with many of the other results in this paper, were presented at a seminar at Jussieu in Paris in June 1997. A similar derivation was given by Bouchaud and Cont (1998), who acknowledge their attendance in a footnote. Although this derivation was used in the 1994 version of this paper, in the 1998 version I gave another derivation based on a "no arbitrage" condition. This turned out to be wrong. In fact, this is sufficient to show that $\phi$ must be increasing, but not to determine its form.

For an equilibrium model the clearing price depends only on the current demand functions. For a general nonlinear price formation rule, in contrast, the price at any time depends on the full sequence of previous net orders. The log-linear rule is somewhere in between: The price change over any given period of time depends only on the net order imbalance during that time. In fact, we can make this a requirement, and use it to derive the log-linear rule: Suppose we require that two orders placed in succession result in the same price as a single order equal to their sum, i.e.

$$f(f(P, \omega_1), \omega_2) = f(P, \omega_1 + \omega_2).$$ (Eq 6)

By grouping orders pairwise, repeated application of equation (6) makes it clear that the price change in any time interval only depends on the sum of the net orders in this interval. Substituting equation (4) into equation (6) gives

$$\phi(\omega_1 + \omega_2) = \phi(\omega_1)\phi(\omega_2).$$

This functional equation for $\phi$ has the solution

$$\phi(\omega) = e^{\omega/\lambda},$$ (Eq 7)

which is equivalent to equation (5). Other possible solutions are $\phi(\omega) = 0$ and $\phi(\omega) = 1$, but neither of these satisfy the requirement that $\phi$ is increasing.

The log-linear price formation rule is only an approximation. It is, however, perhaps the simplest one that gives reasonable results. In addition to the path-independence derived above, it has several other special properties, as shown in Section 4. It simplifies the attribution of profits by making it possible to decompose them pairwise based on correlations of positions. It also implies that total realized wealth is conserved in closed systems. The demonstration of these properties is interesting in part because it makes it clear how nonlinearities lead to path dependence, non-decomposability, and non-conservation of realized wealth.

### 2.2.3  Problems with these assumptions

The derivation above implicitly assumes that market impact is permanent. That is, price changes caused by a net order at any given time persist until new net orders cause other changes). In contrast, if the market impact is temporary, price changes decay, even without new order flow.

The assumption that the market impact function $\phi$ depends only on the net order $\omega$ does not take into account the market maker's risk aversion. Real market makers use their ability to manipulate the price to keep their positions as small as possible. Consideration of such inventory effects makes the price formation process depend on the market maker's position, which depends on past as well as present orders[11]. The assumption that the new price depends only on the most recent order and price also neglects other possible influences. For example, news might change the price directly, without any intervening order flow. Such possibilities are crudely modeled here by adding an external noise term (see equation (8)).

---

11. For an example of an empirical price formation rule that depends on the market maker's position see, e.g. Huang and Stoll (1994).

There is an implicit assumption that the market is symmetric in the sense that there is no *a priori* difference between buying and selling. Indeed, with any price formation rule that satisfies equation (6) buying and selling are inverse operations. (This is clear by letting $\omega_2 = -\omega_1$, which implies that $f^{-1}(P, \omega) = f(P, -\omega)$). This is reasonable for currency markets and many derivative markets, but probably not for most stock markets. The short selling rules in the American stock market, for example, make one expect that the market impact of buying and selling are different. This was observed by Chan and Lakonishok (1993, 1995). Such asymmetries can be taken into account by letting the liquidity for buying and selling be different.

I wish to emphasize that I do not consider the log linear rule to be an accurate prediction of the true market impact function $\phi$. Indeed, several different empirical studies suggest that the shift in the logarithm of the price shift plotted against order size is a concave nonlinear function[12]. The derivation presented here merely justifies the log-linear market impact function as a reasonable starting approximation. Its simplicity facilitates analytic calculations that can be useful in gaining insight into other issues.

## 2.3 Dynamics

We can now write down a dynamical system describing the interaction between trading decisions and prices. Letting $p_t = \log P_t$, and adding noise, equation (5) becomes

$$p_{t+1} = p_t + \frac{1}{\lambda} \sum_{i=1}^{N} \omega^{(i)}(p_t, p_{t-1}, ..., I_t) + \xi_{t+1}.$$

(Eq 8)

I have added a random term $\xi_t$ to account for possible external perturbations to the price that are not driven by trading, such as news announcement or perceived arbitrage possibilities in related markets. The dynamics of equation 8 are completely general. Depending on the collection of functions $\omega^{(i)}$ they can have stable fixed points, limit cycles, or chaotic attractors, or they can be globally unstable. The function $\omega^{(i)}$ is defined in terms of the positions by equation (2).

There is some arbitrariness in what is meant by a "strategy". If an agent switches randomly between two strategies, their combination can be regarded as a single mixed strategy. In terms of the price dynamics, $N$ agents all using the same strategy $x^{(i)}$ are equivalent to a single agent with strategy $Nx^{(i)}$. However, when it comes to selection, as discussed in Section 4.4.2, these are not necessarily equivalent. The superscript ($i$) can either refer to a given agent or a given strategy, depending on the context.

---

12. For discussions of empirical evidence concerning market impact see Hausman and Lo (1992), Chan and Lakonishok (1993, 1995), Campbell et al. (1997), Torre (1997), and Keim and Madhaven (1999). Zhang (1999) has offered a heuristic derivation of a nonlinear market impact rule.

The choice of a discrete time, synchronized trading process is a matter of convenience. We could alternatively have used an asynchronous process with random updating (which is also easy to simulate), or a continuous time Weiner process (which has advantages for obtaining analytic results). The modification of equation (8) is straightforward in either case. The time $\Delta t$ corresponding to a single iteration should be thought of as the timescale on which the fastest traders observe and react to the price, e.g. a minute to a day.

There are several free parameters in equation (8). These include the magnitude of $\xi$, the liquidity $\lambda$, $\Delta t$, and the scale of $\omega^{(i)}$. The last three are not independent. To see this, suppose we make a change of scale, $\omega^{(i)} \rightarrow c^{(i)}\omega^{(i)}$, where $c^{(i)} > 0$. The scale parameter $c^{(i)}$ is proportional to the *capital* of trader $i$, and controls the size of his orders and positions. The dynamics of equation (8) depend only on the non-dimensional ratios $\alpha^{(i)} = c^{(i)}/\lambda$, so doubling the liquidity is equivalent to doubling the scale of all the strategies. Similarly, in the limit where the iteration interval $\Delta t \rightarrow 0$, increasing $\Delta t$ is equivalent to increasing $\lambda$ by the same factor.

There is also a question of units. $x$, $\omega$ and $\lambda$ can be converted from units of shares to units of money by multiplying by the price $P_t$. We can use either set of units based on convenience. For the results on dynamics in Section 3 this distinction is irrelevant, but many of the results concerning profits in Section 4 depend on measuring positions, orders, and liquidity in terms of money.

## 2.4  Comparison to other methods of price formation

### 2.4.1  Temporary equilibrium and market clearing

To compare to a standard temporary equilibrium market clearing model for price formation (e.g. Grossman and Stiglitz 1980, Grandmont 1988, or Arthur et al. 1997), write the total demand $D$ as a function of the logarithm of the price $p$. Suppose $D(p)$ changes by a small amount $\delta D$, i.e. $D(p) \rightarrow D(p) + \delta D(p)$. For the market to clear the total demand must remain constant. To first order in $\delta D$ this implies that $\delta D(p) + D'(p)\delta p = 0$, or

$$\delta p = \frac{-\delta D}{D'(p)}. \tag{Eq 9}$$

It is tempting to identify $\omega = \delta D$ and $\lambda = -D'(p)$, which makes equation (5) and equation (9) the same.

However, there are several important differences. Equation (8) is more realistic in that the price change at time $t + 1$ depends on decisions made based on information available at time $t$. As a result demands are generally not satisfied, and demand and position are not equivalent. In addition, the formalism here makes it clear that there is coordination problem that must be solved to achieve equilibrium: From the point of view of the market maker the strategies of directional traders are uncertain, and from the point of view of the directional traders the liquidity is uncertain. Yet to satisfy equation (9), the liquidity must be the derivative of the agents' total demand. For the equilibrium approximation to be reasonable each agent must learn enough about the other to maintain this coordination, despite ongoing variations in both liquidity and capital.

Non-equilibrium effects are likely to be more important on shorter timescales, where the lack of market clearing is important. Equilibrium models should be more accurate on longer timescales, e.g. a month or more, where non-clearing effects have decayed, and where errors due to modeling with the wrong market impact function might accumulate.

Perhaps the biggest advantage of the non-equilibrium model developed here is simplicity[13]. Finding a clearing price in a temporary equilibrium model requires a cumbersome search over prices. Even in the simplest case where the individual demand functions are linear, this involves solving a system of simultaneous equations at each timestep. As a result, analytic results are typically obtainable only for a few time periods. In contrast, here we have a difference equation, giving an infinite period model.

### 2.4.2 Related work in market making

The literature on market making is reviewed by O'Hara (1995). The standard view is that market making is driven by order processing costs, adverse information, and inventory effects. Order processing costs are simply the charges incurred for handling transactions; adverse information occurs because directional traders may possess additional information that tends to reduce market maker profits; inventory effects occur because of the market maker's aversion to risk and her consequent desire to keep her net position as low as possible. Under the model derived here the price is manipulated in the direction of the net of incoming orders. Thus this model includes both an adverse information and an inventory effect (although the inventory effect is weak, since it does not depend on the market maker's position).

The most relevant point of comparison is Kyle's (1985) model for continuous auctions and insider trading. There are three types of agents: market makers, an insider with perfect knowledge of the liquidation value of the asset (the eventual final price), and noise traders who submit random orders. The final liquidation value and the orders of the noise traders are both normally distributed. By imposing a temporary equilibrium he derives a price formation rule of the form

$$\tilde{P} - P = \omega / \lambda. \qquad \text{(Eq 10)}$$

---

13. Treating the orders of all agents as equivalent greatly simplifies matters. In an equilibrium model agents with steeper demand functions exert more influence on the price. If we were to construct an analogous market impact function $\hat{\phi}$ corresponding to equilibrium price formation, it would depend on the entire set of $N$ orders. For the market making model used here, in contrast, $\phi$ has a one dimensional domain. Thus out-of-equilibrium price formation in this context is simpler and more highly constrained than in equilibrium.

This result comes from maximizing the profits of the insider while otherwise assuming market efficiency (no profits for the market maker). A similar price formation rule was also derived by Grossman and Stiglitz (1980). It is encouraging that these models give qualitatively similar results, even though the assumptions are so different.

There is an important difference, however: Equation (10) is stated in terms of the price, whereas equation (5) is stated in terms of the logarithm of the price. Kyle's price formation rule fails to satisfy the obvious boundary condition that the price remains positive. This problem can be traced back to the assumption of normally distributed final liquidation values.

An important advantage of Kyle's model is that he is able to derive the liquidity, which is

$$\lambda = 2\sigma_i/\sigma_n, \tag{Eq 11}$$

where $\sigma_i$ is the standard deviation of liquidation values known to the insider, and $\sigma_n$ is the standard deviation of orders submitted by the noise traders. Kyle's results are important in the historical context of the market microstructure literature because they give insight into why informed agents may trade incrementally in order to reveal their true intentions gradually. The impact of a large market order can easily overwhelm what might otherwise be a splendid profit-making opportunity. This is well known to competent market practitioners, who spend considerable effort on *order tactics* to minimize market friction. This has been the subject of recent papers by Bertismas and Lo (1998) and Almgren and Chriss (1999). For a discussion of market friction see Section 4.1.4.

### 2.4.3 Market making with limit orders

In real markets limit orders are an important component of market making. Most market makers keep a *limit order book,* containing unfilled limit orders at each price level, ordered in priority according to their time of reception. A market order can be filled either by transacting with a limit order placed by another trader, or by transacting directly with the market maker. Market making and limit orders go hand-in-hand. Indeed, it is possible to act as a market maker by simultaneously submitting buy and sell limit orders as needed to bracket the current price.

Understanding the role of limit orders in price formation is beyond the scope of this paper[14]. However, it is worth pointing out that when market making is done purely by limit orders, there is a correspondence between the density of limit orders and the market impact function. As before, let prices be continuous, and let the density of limit orders be $\delta(P)$. When a market order $\omega$ is received it is crossed with unfilled limit orders that are of opposite sign, beginning with those closest to the current price. The resulting shift in the price can be found from the condition

---

14. See Bak et al. (1997) and Eliezer and Kogan (1998).

$$\omega = \int_{P_t}^{P_{t+1}} \delta(P')dP'. \qquad \text{(Eq 12)}$$

The density function $\delta(P) = \lambda/P$ is a natural choice, since it is positive and scale independent. Substituting this into equation (12) and doing the integral gives Equation (7). In steady state the density of limit orders will be determined as a balance between the volume of limit orders accumulating in the book, and the volume of market orders and crossing limit orders removing them from the book. We would expect the density $\delta(P)$ and therefore $\lambda$ to increase with the volume of limit orders. Note that Kyle's model predicts that market orders have the opposite effect[15].

---

15. $\sigma_n$ in equation (11) is the standard deviation of random trading, which is proportional to volume. Thus liquidity drops as market order volume goes up. This makes sense: limit orders that do not cross the current price level enhance liquidity, and market orders and limit orders that cross the current price level deplete it. However, when one takes competitive market making into account this is not so obvious: with more market order trading volume there are more opportunities to cross orders and less risk for a given level of profits.

# 3. Ecology

This section introduces several commonly used trading strategies and studies their price dynamics. Since the market maker is in a sense a "neutral" agent, we can begin by studying strategies individually. Each strategy has a characteristic *induced market dynamics,* which provides useful intuition when many strategies are present at once. Throughout this section we will assume that the strategies are fixed for a given simulation. Profits are not reinvested. The consequences of reinvestment and other forms of capital allocation are studied in Section 4.

There are two reasons for titling this chapter "Ecology". The first is that, in analogy with biology, it stresses the interrelationships of financial agents with each other and their environment. The interactions of financial agents are strongly mediated through a single variable, the price, which forms an important part of their environment. Each trading strategy influences the price, and the price in turn influences each trading strategy.

The second reason has to do with the approach. An ecologist studying the interaction of mountain lions and deer begins by describing their behavior, without worrying about why it occurs. Similarly, I will simply take some of the most common trading strategies as given and study their price dynamics. Of course, explaining why these strategies are used in terms of a broader principle, such as utility maximization, is a very interesting question -- but it is beyond the scope of this paper.

Despite the wide variation in financial trading strategies, we can classify them into broad groups. One method of classification is based on information inputs. Decision rules that depend only on the price history are called *technical* or *chartist* strategies. *Trend following* strategies are a common special case in which positions are positively correlated with past price movements. *Value* or *fundamental* strategies are based on a subjective assessment of value in relation to price.

We will show that over short timescales value investing strategies induce negative autocorrelations in prices. Real markets have autocorrelations close to zero, and so apparently do not consist purely of value investors. Trend following strategies, in contrast, induce positive short term autocorrelations. By making an appropriate combination of value strategies and trend-following strategies, the price series can have low autocorrelations, even though it has strong nonlinear structure that is nonetheless difficult to detect. A combination of value investors and trend followers gives rise to commonly observed market phenomena such as fat tails in the distribution of log-returns, correlated volume and volatility, and temporal oscillations in the difference between prices and values.

Induced autocorrelations on longer timescales are generally complicated. Strategies are necessarily formulated in terms of positions, which determine both risk and profits. Market impact, however, is caused by orders, which are the time difference or derivative of positions. Thus the non-equilibrium price dynamics induced by any given strategy generally have second order oscillatory terms, as demonstrated in the following examples.

## 3.1  Value investors

Value investors make a subjective assessment of value in relation to price. They believe that value is not fully reflected in the price: If an asset is undervalued the price will tend to rise, and if it is overvalued the price will tend to fall. They attempt to make profits by taking positive (long) positions when they think the market is undervalued and negative (short) positions when they think the market is overvalued.

Value is inherently subjective. The assessment of value might involve the analysis of fundamental data, such as earnings and sales, or judgements based on the quality of management. For the purposes of this paper I don't care how individual agents form their opinions about value. We take this as given. The focus is rather on the trading strategy, which can lead to interesting dynamics in the price and its relation to value.

A market can be viewed as an organ of society that performs the function of resource allocation. Markets help set society's goals. If the price of pork bellies go up, people will grow more pigs. One measure of how well markets perform their function is whether prices correspond to a consensus view of values. Of course perceptions of value differ -- this drives trading. Nonetheless, if markets perform their function efficiently, prices should track aggregate values, at least over the long term. There is no reason to assume that this happens *a priori* -- it should be possible to derive it from first principles. Evidence from market data suggests that, while prices track values over the very long term, large deviations are the rule rather than the exception. Campbell and Shiller (1988) call this *excess volatility*. We will see that this arises naturally from the price dynamics of value strategies, particularly when there are nonlinearities and diverse views.

Earlier studies of nonequilibrium price formation (e.g. Fisher 1983), assume that value is fixed. This oversimplifies the problem, tilting the analysis in favor of stability. Here we will assume that the perceived value is given by an exogenous random process. Let the logarithm of the perceived value $v_t$ be a random walk,

$$v_{t+1} = v_t + \eta_{t+1}, \qquad \qquad \text{(Eq 13)}$$

where $\eta_t$ is a normal, IID noise process with standard deviation $\sigma_\eta$ and mean $\mu_\eta$. We will begin by studying the case where everyone perceives the same value, and return to study the case where there are diverse perceived values in Section 3.1.5.

The natural way to quantify whether price tracks value is by using the concept of cointegration, introduced by Engle and Granger (1987). This concept is motivated by the possibility that two random processes can each be random walks, even though on average they tend to move together and stay near each other. More specifically, two random processes $y_t$ and $z_t$ are *cointegrated* if there is a linear combination $u_t = ay_t + bz_t$ that is stationary. For example, price and value are cointegrated if $p_t - v_t$ has a well defined mean and standard deviation.

### 3.1.1 Simple value strategies

For the simplest class of value strategies the position is of the form

$$x_{t+1} = x(v_t, p_t) = V(p_t - v_t),$$  (Eq 14)

where $V$ is a generally decreasing function with $V(0) = 0$, $v_t$ is the logarithm of the perceived value, and $p_t$ is the logarithm of the price. *Generally decreasing* means that $V$ either decreases or remains constant, and is not constant everywhere[15]. This class of strategies only depends on the *mispricing $m_t = p_t - v_t$*. Such a strategy takes a positive (long) position when the asset is underpriced; if the asset becomes even more underpriced, the position either stays the same or gets larger. Similarly, if the mispricing is positive it takes a negative (short) position.

If $V$ is differentiable we can expand it in a Taylor series. To first order the position can be approximated as

$$x_{t+1} = -c(p_t - v_t),$$

where $c > 0$ is a constant proportional to the trading capital. From equations (2) and (8) the induced price dynamics in a market consisting only of this strategy and the market maker are

$$\Delta p_{t+1} = \frac{c}{\lambda}(\Delta v_t - \Delta p_t) + \xi_{t+1},$$  (Eq 15)

where $\Delta$ is the time difference operator, e.g. $\Delta p_{t+1} = p_{t+1} - p_t$. Letting $r_t = \Delta p_t$, $\alpha = c/\lambda$, and $\eta_t = \Delta v_t$, this can be written

$$\begin{aligned} r_{t+1} &= -\alpha r_t + \alpha \eta_t + \xi_{t+1} \\ p_{t+1} &= p_t + r_{t+1} \end{aligned}.$$  (Eq 16)

These dynamics are second order. This is evident from equation (15) since $p_{t+1}$ depends on both $p_t$ and $p_{t-1}$. The stability can be determined by neglecting the noise terms and writing equation (16) in the form $u_{t+1} = Au_t$, where $u_t = (r_t, p_t)$. The eigenvalues of $A$ are $(1, -\alpha)$. Thus when $\alpha \leq 1$ the dynamics are neutrally stable, which implies that the logarithm of the price, like the logarithm of the value, follows a random walk. When $\alpha > 1$ the dynamics are unstable.

Simple value strategies induce negative first autocorrelations in the log-returns $r_t$. This is easily seen by multiplying both sides of equation (16) by $r_{t-i}$, subtracting the mean, and taking time averages. Let

---

15. This also allows for the possibility of positions of a single sign, e.g. only long positions.

$$\rho_r(\tau) \ = \ \frac{\langle r_t r_{t-\tau}\rangle - \langle r_t \rangle^2}{\sigma_r^2}, \tag{Eq 17}$$

where $\sigma_r$ is the standard deviation of $r$. Assuming stationarity, this gives the recursion relation $\rho_r(\tau) = -\alpha\rho_r(\tau-1)$. Since $\rho_r(0) = 1$, this implies

$$\rho_r(\tau) \ = \ (-\alpha)^\tau, \tag{Eq 18}$$

where $\tau = 0, 1, 2, \dots$. Because $\alpha > 0$, the first autocorrelation is always negative. Since the autocorrelation is determined by the linear part of $V$, this is true for any differentiable value strategy in the form of equation (14).

The trading of this strategy amplifies noise. To see this, compute the variance of the log-returns by squaring equation (15) and taking time averages. This gives

$$\sigma_r^2 \ = \ \frac{\alpha^2\sigma_\eta^2 + \sigma_\xi^2}{1-\alpha^2}, \tag{Eq 19}$$

where $\sigma_\eta^2$ and $\sigma_\xi^2$ are the variances of $\eta_t$ and $\xi_t$. This amplifies the external noise, since for any value of $\alpha$, $\sigma_r > \sigma_\xi$. Similarly, if $\alpha > 1/\sqrt{2}$ then $\sigma_r > \sigma_\eta$. In this case the volatility in the log-returns is greater than that in the log-values, and it amplifies the noise in the value process.

The most surprising result is that prices do not track values. This is evident because equation (16) shows no explicit dependence on price or value. The lack of cointegration can be shown explicitly by substituting $m_t = p_t - v_t$ into equation (16), which gives

$$\Delta m_{t+1} \ = \ -\alpha\Delta m_t - \eta_t + \xi_t.$$

When $\alpha < 1$, $\Delta m_t$ is stationary and $m_t$ is a random walk. Numerical simulations and heuristic arguments also support this when $V$ is nonlinear[16]. The intuitive reason for this is that, while a trade entering a position moves the price toward value, an exiting trade moves it away from value. Thus, while the negative autocorrelation induced by simple value strategies might reduce the rate at which prices drift from value, this is not sufficient for cointegration. This is illustrated in Figure 1.

The lack of cointegration can lead to problems with unbounded positions, implying unbounded risk. This comes about because the mispricing is unbounded, and the position is proportional to the mispricing. Thus if this is the only strategy present in the market the position is also unbounded. This problem disappears if another strategy is present in the market that cointegrates prices and values.

---

16. In the general nonlinear case the mispricing can be written
$\Delta m_{t+1} = 1/\lambda(V(m_{t+1}) - V(m_t)) - \eta_t + \xi_t$. Because $V$ is generally decreasing, this can be written $\Delta m_{t+1} = -c(m_t)\Delta m_t - \eta_t + \xi_t$, where $c(m_t) \geq 0$. It seems that either this is a stable random process, in which case $m_t$ is a random walk, or it is unstable, in which case $m_t$ is unstable as well.

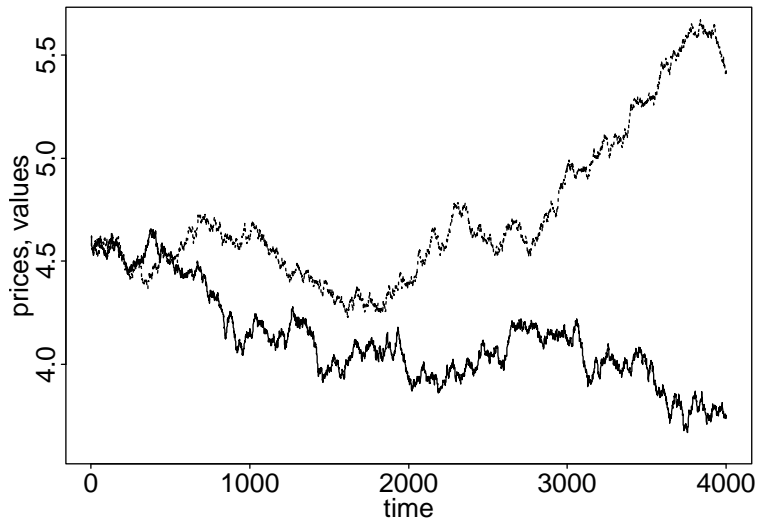

**FIGURE 1. The log-price $z_t$ (dashed line) and the log-value $v_t$ (solid line) for the linear position based value strategy of equation (16) with $\alpha = 0.1$. While changes in the price are negatively correlated, there is no cointegration, and price fails to track value.**

The conclusion that prices and values are not cointegrated holds for any $\alpha$. Thus this problem doesn't disappear under arguments of market efficiency (which might adjust $\alpha$). For prices to track values some other class of strategies must be present, e.g. the more complicated value strategies discussed in Section 3.1.3.

So far we have assumed ongoing changes in value. It is perhaps even more surprising that the price fails to converge even if the value changes once and then remains constant. To see this, consider equation (15) with $\Delta v_1 = v$, and $\Delta v_t = 0$ for $t > 1$. Assume $\xi_t = 0$, and for convenience let $p_1 = v_1 = 0$ and $\Delta p_1 = 0$. Iterating a few steps by hand shows that $p_t = (\alpha - \alpha^2 + \alpha^3 + \dots (-\alpha)^{t-1})v$. If $\alpha < 1$, in the limit $t \to \infty$ this converges to $p_\infty = \alpha v / (1 + \alpha)$. Thus when $\alpha < 1$ the price initially moves toward the new value, but it never reaches it; when $\alpha > 1$ the dynamics are unstable.

The intuitive reason for this is that the strategy is formulated in terms of position. If the value suddenly increases, this strategy buys. But if the difference between price and value persist at the same level, it maintains a constant position. Mispricings can persist forever. The next section considers a more aggressive strategy that buys until prices match values.

### 3.1.2 Order-based value strategies

One way to make prices track values is to make the strategy depend on the order instead of the position. Under the simple value strategy above, if the mispricing reaches a given level, the trader takes a position. If the mispricing holds that level, he keeps the same position. For an order-based strategy, in contrast, if the asset is underpriced he will buy, and he will keep buying as long as it remains underpriced. One can define an *order based value strategy* of the form

$$\omega_{t+1} = \omega(v_t, p_t) = W(p_t - v_t)$$

where as before $W$ is a generally decreasing function with $W(0) = 0$. If we again expand in a Taylor's series, then to leading order this becomes

$$\omega_{t+1} = -c(p_t - v_t).$$

Without presenting the details, let me simply state that it is possible to analyze the dynamics of this strategy and show that the mispricing has a well-defined standard deviation. Prices track values. The problem is that the position is unbounded. This is not surprising, given that this is such an odd and unrealistic strategy. The trader keeps buying as long as the asset is underpriced, and by the time the mispricing goes to zero, the position may be arbitrarily large. Furthermore, the problem of unbounded positions occurs even in the presence of other strategies that cause cointegration of price and value. Numerical experiments suggest that non-linear extensions have similar problems.

It is unrealistic to consider strategies that do not have bounded positions. Real traders have risk constraints, which mean that their positions always have limits. I have studied hybrid strategies, in which orders depend on the mispricing with a bound on positions. Prices may track values for a period of time, but as soon as the bound is reached, the price escapes. Order-based strategies do not offer a realistic solution to the problem of prices tracking values.

### 3.1.3  State-dependent threshold value strategies

The analysis above poses the question of whether there exist strategies that cointegrate prices and values and have bounded risk at the same time. This section introduces a class of strategies with this property.

From the point of view of a practitioner, a concern with the simple position-based value strategies of Section 3.1.1 is excessive transaction costs. Trades are made whenever the mispricing changes. As shown in Section 4.1.4, random trading tends to induce losses through market friction. A common approach to ameliorate this problem and reduce trading frequency is to use state dependent strategies, with a threshold for entering a position, and another threshold for exiting it. Like the simpler value strategies studied earlier, such strategies are based on the belief that the price will revert to the value. By only entering a position when the mispricing is large, and only exiting when it is small, the goal is to trade only when the expected price movement is large enough to beat transaction costs.

An example of such a strategy, which is both nonlinear and state dependent, can be constructed as follows: Assume that a short position $-c$ is entered when the mispricing exceeds a threshold $T$ and exited when it goes below a threshold $\tau$. Similarly, a long position $c$ is entered when the mispricing drops below a threshold $-T$ and exited when it exceeds $-\tau$. This is illustrated in Figure 2. Since this strategy depends on its own position as well as the mispricing, it can be thought of as a finite state machine, as shown in Figure 3.

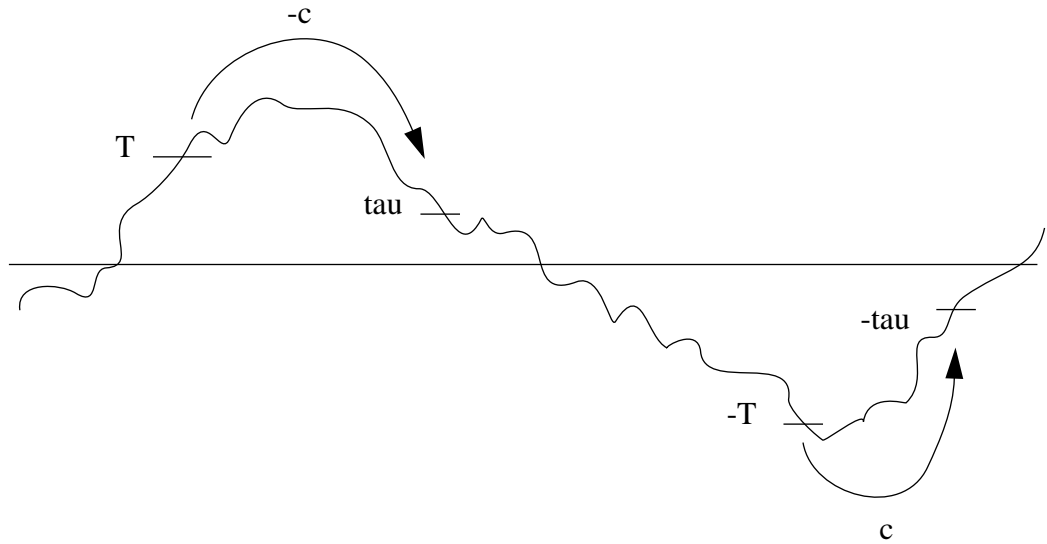

**FIGURE 2. Schematic view of a nonlinear, state-dependent value strategy. The trader enters a short position $-c$ when the mispricing $m_t = p_t - v_t$ exceeds a threshold $T$, and holds it until the mispricing goes below $\tau$. The reverse is true for long positions.**

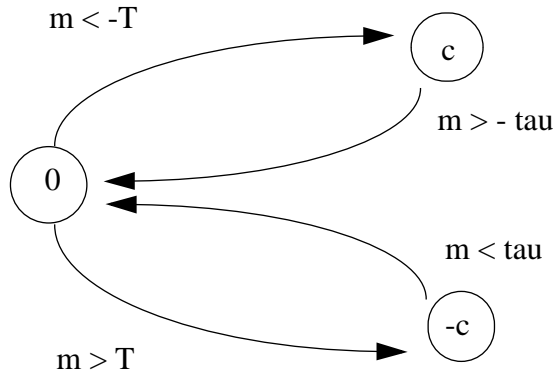

**FIGURE 3. The nonlinear state-dependent value strategy represented as a finite-state machine. From a zero position a long-position $c$ is entered when the mispricing $m$ drops below the threshold $-T$. This position is exited when the mispricing exceeds a threshold $-\tau$. Similarly, a short position $-c$ is entered when the mispricing exceeds a threshold $T$ and exited when it drops below a threshold $\tau$**

In general different traders will choose different entry and exit thresholds. Let trader $i$ have entry threshold $T^{(i)}$ and exit threshold $\tau^{(i)}$. For the simulations presented here we will assume a uniform distribution of entry thresholds ranging from $T_{min}$ to $T_{max}$, and a uniform density of exit thresholds ranging from $\tau_{min}$ to $\tau_{max}$, with a random pairing of entry and exit thresholds. $c$ is chosen so that $c = a(T - \tau)$, where $a$ is a positive constant[17].

There are several requirements that must be met for this to be a sensible value strategy. The entry threshold should be positive and greater than the exit threshold, i.e. $T > 0$ and $T > \tau$. In contrast, there are plausible reasons to make $\tau$ either positive or negative. A trader who is very conservative about transaction costs, and wants to be sure that the full return has been extracted before the position is exited, will take $\tau < 0$. However, others might decide to exit their positions earlier, under the theory that once the price is near the value there is little expected return remaining. We can simulate a mixture of the two approaches by making $\tau_{min} < 0$ and $\tau_{max} > 0$. However, to be a sensible value strategy, a trader would not exit a position at a mispricing that is further from zero than the entry point. $\tau_{min}$ should not be *too* negative, so we should have $-T < \tau < T$ and $|\tau_{min}| \le T_{min}$.

$\tau < 0$ is a desirable property for cointegration. When this is true the price changes induced by trading always have the opposite sign of the mispricing. This is true both entering and exiting the position. A simulation with $\tau_{max} = 0$ and $\tau_{min} < 0$ is shown in Figure 4. Numerical tests clearly show that the price and value are cointegrated. The coin-tegration is weak, however, in the sense that the mispricing can be large and keep the same sign for many iterations.

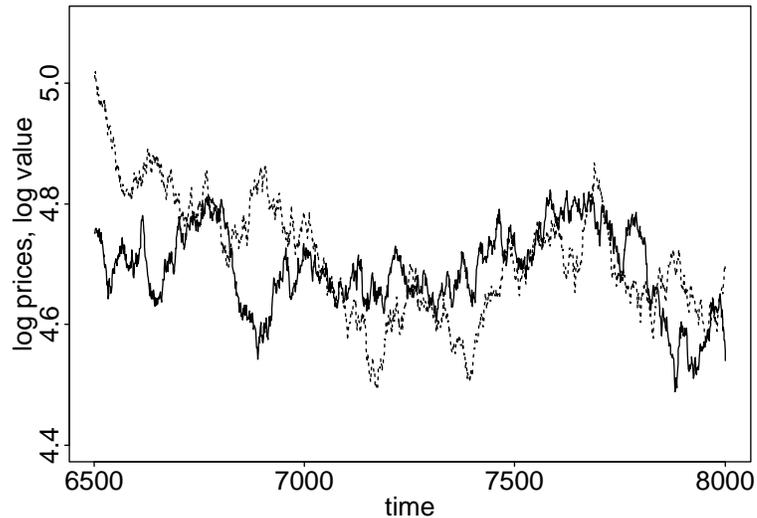

**FIGURE 4. The induced price dynamics of a nonlinear state-dependent value strategy with 1000 traders using different thresholds. The log-price is shown as a solid line and the log-value as a dashed line.** $\tau_{min} = -0.5$, $\tau_{max} = 0$, $T_{min} = 0.5$, $T_{max} = 6$, $N = 1000$, $a = 0.001$, $\sigma_\eta = 0.01$, and $\sigma_\xi = 0.01$, and $\lambda = 1$.

---

17. This assignment is natural because traders managing more money (with larger $c$) incur larger transaction costs. Traders with larger positions need larger mispricings to make a profit.

Figure 5 shows a simulation with the range of exit thresholds chosen so that $\tau_{min} < 0$ but $\tau_{max} > 0$. For comparison with Figure 4 all other parameters are the same. The price

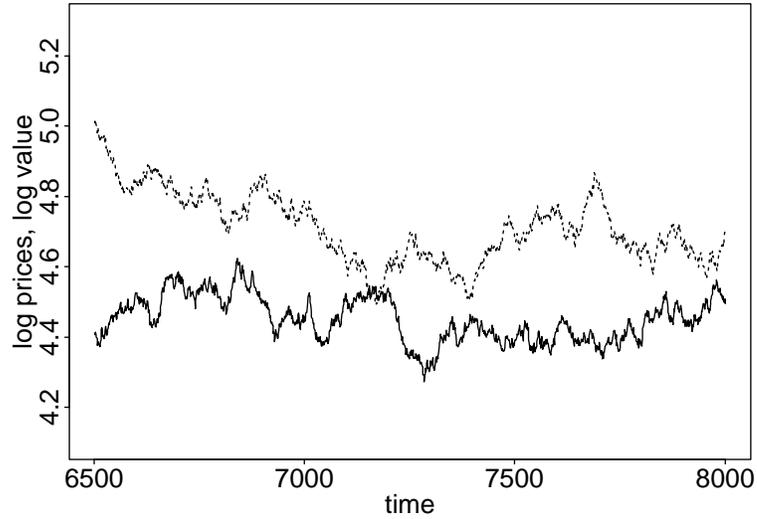

**FIGURE 5. Price (solid) and value (dashed) vs. time for the nonlinear state-dependent strategy of Figure 3. The parameters and random number seed are the same as Figure 4, except that $\tau_{min} = -0.5$ and $\tau_{max} = 0.5$**

and value are still cointegrated, but more weakly than before. This is apparent from the increased amplitude of the mispricing. In addition, there is a tendency for the price to "bounce" as it approaches the value. This is caused by the fact that when the mispricing approaches zero some traders exit their positions, which pushes the price away from the value. The value becomes a "resistance level" for the price (see e.g. Edwards and Magee, 1992), and there is a tendency for the mispricing to cross zero less frequently than it does when $\tau^{(i)} < 0$ for all $i$. Based on results from numerical experiments it appears that the price and value are cointegrated as long as $\tau_{min} < 0$. Necessary and sufficient conditions for cointegration deserve further study[18].

This demonstrates that there is at least one class of value strategies that cointegrates price and value. An encouraging property is that the strength of the cointegration relationship is realistically weak, with mispricings that can persist for thousands of iterations. However, it is surprising that cointegration of price and value should depend on something as indirect as state-dependence induced by the motivation to reduce transaction costs.

---

18. Problems can occur in the simulations if the capital $c = a(T - \tau)$ for each strategy is not assigned reasonably. If $a$ is too small the traders may not provide enough restoring force for the mispricing; once all $N$ traders are committed to a long or short position, price and value cease to be cointegrated. If $a$ is too big instabilities can result because the kick provided by a single trader creates oscillations between entry and exit. Nonetheless, between these two extremes there is a large parameter range with reasonable behavior.

### 3.1.4 Technical enhancements of value strategies

Many traders use technical signals to enhance value strategies. A commonly used class are "value strategies with technical confirmation signals". A trader may believe in value, but also believe that the price history provides information about what others are doing. For instance, one can modify the threshold value strategy by making the entry condition for a short position of the form $m_t > T$ and $p_{max} - p_t > C$, where $p_{max}$ is a running maximum and $C > 0$. The second condition measures a turning point in the price. By taking the logical "and" of these two conditions, the trader hopes to reduce risk, by waiting until the market indicates that other value traders are starting to enter their positions. Thus a pure value strategy may begin a price reversal and a technical enhancement may reinforce it. Simulations show that a mixture of these strategies with the strategies of the previous section make prices track values more closely.

### 3.1.5 Heterogeneous values, representative agents, and excess volatility

So far we have assumed a single perceived value, but given the tendency of people to disagree, in a more realistic setting there will be a spectrum of different values. We will show that in this case, for strategies that are linear in the logarithm of value, the price dynamics can be understood in terms of a single *representative agent,* whose perceived value is the mean of the group. However, for nonlinear strategies this is not true -- there exists no representative agent, and diverse perceptions of value cause excess volatility.

Suppose there are $N$ different traders perceiving value $v_t^{(i)}$, using a value strategy $V^{(i)}(v_t, p_t) = c^{(i)} V(v_t, p_t)$, where $c^{(i)}$ is the capital of each individual strategy. The dynamics are

$$p_{t+1} = p_t + \frac{1}{\lambda} \sum_{i=1}^{N} c^{(i)} V(v_t^{(i)}, p_t),$$

Providing the strategy is linear in the value the dynamics will be equivalent to that of a single agent with the average perceived value and the combined capital. This is true if $V$ satisfies the property

$$\sum_{i=1}^{N} c^{(i)} V(v_t^{(i)}, p_t) = c V(\bar{v}_t, p_t),$$

where

$$\bar{v}_t = \frac{1}{c} \sum_{i=1}^{N} c_i v_t^{(i)},$$

and $c = \sum c_i$. For example, the linearized value strategy of Section 3.1.1 satisfies this property. Thus, for strategies that depend linearly on the logarithm of value, the mean is sufficient to completely determine the price dynamics, and the fact that opinions are diverse is unimportant. The market dynamics are those of a single representative agent.

The situation is quite different when the strategies depend nonlinearly on the value. To demonstrate how this leads to excess volatility, we will study the special case where traders perceive different values, but these values change in tandem. This way we are not introducing any additional noise to the value process by making it diverse, and any amplification in volatility clearly comes from the dynamics rather than something that has been added. The dynamics of the values can be modeled as a simple reference value process $\bar{v}_t$ that follows equation (13), with a fixed random offset $v^{(i)}$ for each trader. The value perceived by the $i^{th}$ trader at time $t$ is

$$v_t^{(i)} = \bar{v}_t + v^{(i)}. \qquad \text{(Eq 20)}$$

In the simulations the value offsets are assigned uniformly between $v_{min}$ and $v_{max}$, where $v_{min} = -v_{max}$, so that range is $2v_{max}$.

We will define the excess volatility as

$$\varepsilon = \sqrt{\sigma_r^2/(\sigma_\eta^2 + \sigma_\xi^2)}, \qquad \text{(Eq 21)}$$

i.e. as the ratio of the volatility of the log-returns to the volatility of the noise terms. This measures the noise amplification. If $\varepsilon > 1$ the log-returns of prices are more volatile than the fluctuations driving the price dynamics. Figure 6 illustrates how the excess volatility increases as the diversity of perceived values increases, using the threshold value strategy of Section 3.1.3. The excess volatility also increases as the capital increases. This is caused by additional trading due to disagreements about value. If the market is a machine whose purpose is to keep the price near the value, this machine is noisy and inefficient.

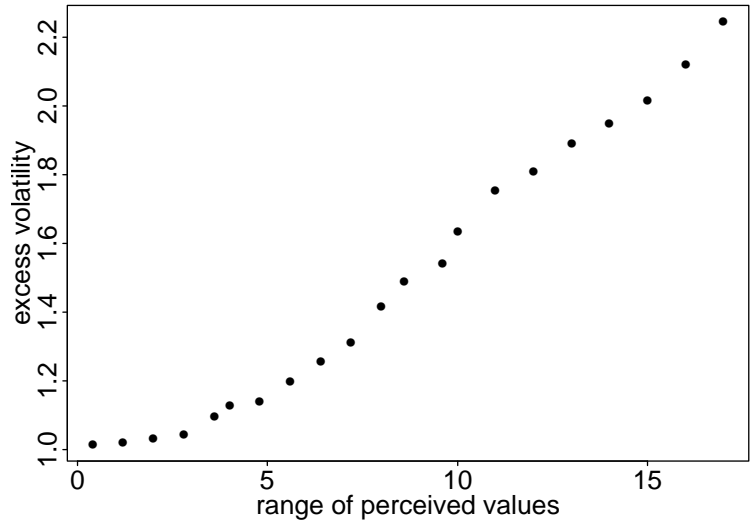

**FIGURE 6. Excess volatility as the range of perceived values increases while the capital is fixed at $0.035$. See equation (21). The other parameters are the same as those in Figure 4.**

### 3.1.6 Influence of price on value?

We have seen that it is difficult to find strategies that make prices track values. For all of the strategies studied here the value is treated as an exogenous external input. Prices react to values, but values do not react to prices. In reality, of course, peoples' perceptions of value are influenced by prices, a phenomenon that Soros (1987) calls *market reciprocity*. Since not all information is common knowledge, the view that prices reflect new and valuable information is very sensible (Grossman 1989). Seen in this context it is rational to adjust one's perception of value based on prices. Future work will investigate this possibility.

## 3.2 Trend followers

*Trend followers,* also sometimes called positive-feedback investors, invest based on the belief that price changes have inertia. A trend strategy has a positive (long) position if prices have recently been going up, and a negative (short) position if they have recently been going down. More precisely, a trading strategy is trend following on timescale $\theta$ if the position $x_t$ has a positive correlation with past price movements on timescale $\theta$, i.e.

$$correlation(x_{t+1}, (p_t - p_{t-\theta})) > 0.$$

A strategy can be trend following on some timescales but not on others.

An example of a simple linear trend following strategy, which can be regarded as a first order Taylor approximation of a general trend following strategy, is

$$x_{t+1} = c(p_t - p_{t-\theta}), \tag{Eq 22}$$

where $c > 0$. Note that if we were to let $c < 0$ this would become a *contrarian strategy*. From equation (8), the induced dynamics are

$$r_{t+1} = \alpha(r_t - r_{t-\theta}) + \xi_t. \tag{Eq 23}$$

where $\alpha = c/\lambda > 1$ and $r_{t-\theta} = p_t - p_{t-\theta}$. Figure 7 shows a time series of prices.

The stability of the dynamics can be calculated by writing equation (23) in the form $u_{t+1} = Au_t$, where $u_t = (r_t, ..., r_{t-\theta})$, and computing the eigenvalues of $A$. For $\theta = 1$ the eigenvalues are

$$\varepsilon_\pm = \frac{\alpha(1-\alpha) \pm \sqrt{5 - 2\alpha + \alpha^2}}{2}.$$

The dynamics are stable when $\alpha < 1$. Note that this is the same stability condition derived for the simple value strategy of equation (16).

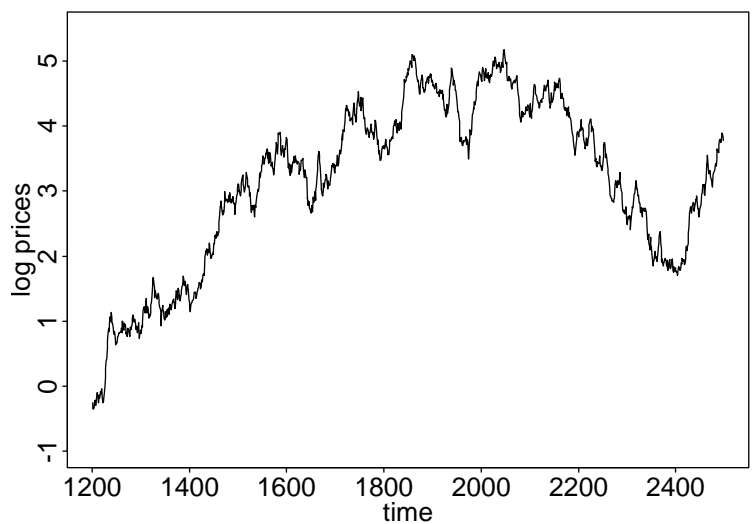

**FIGURE 7. Log price vs. time for trend followers with** $\alpha = 0.2$ **and** $\theta = 10$ **in equation (23). Trend followers tend to induce short term trends in prices, but they also cause oscillations on longer timescales.**

Like value strategies, trend strategies amplify noise. The variance of the log-returns can be computed by taking the variance of both sides of equation (23). This gives

$$\sigma_r^2 = \frac{\sigma_\xi^2}{(1 - 2\alpha^2(1 - \rho_r(\theta)))}.$$

Since $\rho_r(\theta) \leq 1$, it follows that $\sigma_r > \sigma_\xi$. Regardless of the value of $\alpha$ or $\rho_r$, the variance of the price fluctuations is larger than that of the noise driving term. However, note that this is also true for a contrarian strategy: Reversing the sign of $c$ in equation (22) leaves this result unchanged. Thus we see that the commonly heard statement that "trend following is destabilizing" is misleading, since contrarian strategies are just as destabilizing.

Trend strategies induce trends in the price, but as we show below, they also induce oscillations of equal intensity on longer timescales. For example, consider Figure 8, which shows the autocorrelation function for Figure 7. The decaying oscillations between positive and negative values are characteristic of trend strategies with large lags. For $\tau = 1$ the autocorrelation function is of order $\alpha$. As $\tau$ increases it decays, crossing zero at roughly $\tau \approx \theta/2 + 1$. As $\tau$ continues to increase it becomes negative, reaching a minimum at $\tau = \theta + 1$, where it is of order $-\alpha$. The autocorrelation then increases again, reaching a local maximum at $\tau = 2\theta + 2$, where it is of order $\alpha^2$. As $\tau$ increases still further it oscillates between positive and negative values with period $2\theta + 2$, decaying by a factor of $\alpha$ with every successive period.

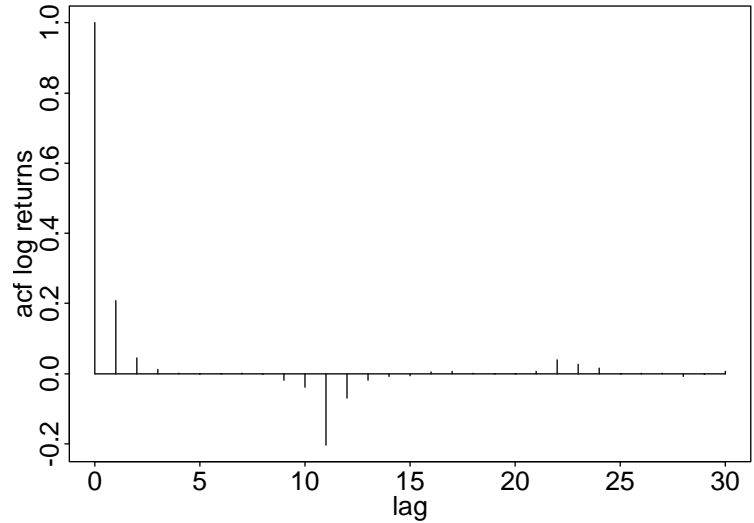

**FIGURE 8. The autocorrelation function for equation (23) with $\alpha = 0.2$ and $\theta = 10$. The complicated structure of the autocorrelation function causes short term trends, with longer term oscillations.**

This behavior can be understood analytically. A recursion relation for the autocorrelation function can be obtained by multiplying equation (23) by $r_{t-n}$, subtracting the mean, and averaging, which gives

$$\rho_r(n+1) = \alpha(\rho_r(n) - \rho_r(|n-\theta|)) . \qquad \text{(Eq 24)}$$

Doing this for $n = 0, \ldots, \theta - 1$ gives a system of $\theta$ linear equations that can be solved for the first $\theta$ values of $\rho_r(\tau)$ by making use of the requirement that $\rho_r(0) = 1$. The remainder of the terms can be found by iteration. For example, for $\theta = 1$, for $\tau = 1, \ldots, 6$ the autocorrelation function is

$$\rho_r(\tau) = \frac{1}{1+\alpha}(\alpha, -\alpha, -2\alpha^2, \alpha^2(1-2\alpha), \alpha^3(3-2\alpha), \alpha^4(3-2\alpha)) . \qquad \text{(Eq 25)}$$

Solving this for a few other values of $\theta$ demonstrates that the first autocorrelation $\rho_r(1)$ is always positive and of order $\alpha$, but $\rho_r(\theta+1)$ is always negative of order $-\alpha$. For large $\theta$ and small $\alpha$, using equation (24) it is easy to demonstrate that the autocorrelation follows the behavior described above. For $\tau \le \theta + 1$, to leading order in $\alpha$, $\rho(\tau) \approx \alpha^\tau - \alpha^{|\tau-\theta-1|+1}$.

The main point is to illustrate that the patterns trend strategies induce in prices are not as simple as one might have thought. There are both positive and negative autocorrelations, and they are roughly equal in magnitude. While trend strategies induce trends at short timescales, they also induce oscillations at longer timescales. This is a consequence of the second order terms that are always present in the dynamics because strategies are formulated in terms of positions rather than orders.

Like value investors, trend followers often use thresholds to reduce transaction costs. Given a trend indicator $I(p_t, ..., p_{t-\theta})$, a nonlinear trend strategy can be defined as a finite state machine, in a similar vein to the value strategy of Section 3.1.3, with thresholds both for entering and existing positions. Strategies of this type will be used in the next section.

## 3.3 Value investors and trend followers together

In this section we[19] investigate the dynamics of trend followers and value investors together, using the threshold value strategies described in Section 3.1.3 and the threshold trend strategies mentioned at the end of Section 3.2. We make a qualitative comparison to annual prices and dividends for the S&P index[20] from 1889 to 1984, using the average dividend as a crude measure of value, and simulating the price dynamics on a daily timescale. As a proxy for daily value data we linearly interpolate the annual logarithm of the dividends, creating 250 surrogate trading days for each year of data. These provide the reference value process $\bar{v}_t$ in equation (20).

The parameters for the simulation are given in Table 2. There were two main criteria for choosing parameters: First, we wanted to match the empirical fact that the correlation of the log-returns is close to zero. This was done by matching the population of trend followers and value investors, so that the positive correlation induced by the trend followers is cancelled by the negative correlation of the value investors. Thus the common parameters for trend followers and value investors are the same. Second, we wanted to match the volatility of prices with the real data. This is done primarily by the choice of $a$ and $N$ in relation to $\lambda$, and secondarily by the choice of $v_{min}$ and $v_{max}$. Finally, we chose what we thought was a plausible timescale for trend following ranging from $1 - 100$ days.

**TABLE 2. Parameters for the simulation with trend followers and value investors in Figure 10.**

| Description of parameter | symbol | value |
|---|---|---|
| number of agents | $N_{value}, N_{trend}$ | 1200 |
| minimum threshold for entering positions | $T_{min}^{value}, T_{min}^{trend}$ | 0.2 |
| maximum threshold for entering positions | $T_{max}^{value}, T_{max}^{trend}$ | 4 |
| minimum threshold for exiting positions | $\tau_{min}^{value}, \tau_{min}^{trend}$ | –0.2 |
| maximum threshold for exiting positions | $\tau_{max}^{value}, \tau_{max}^{trend}$ | 0 |
| scale parameter for capital assignment | $a_{value}, a_{trend}$ | $2.5 \times 10^{-3}$ |
| minimum offset for log of perceived value | $v_{min}$ | –2 |
| maximum offset for log of perceived value | $v_{max}$ | 2 |
| minimum time delay for trend followers | $\theta_{min}$ | 1 |
| maximum time delay for trend followers | $\theta_{max}$ | 100 |
| noise driving price formation process | $\sigma_{\xi}$ | 0.35 |
| liquidity | $\lambda$ | 1 |

---

19. The work in this section was done in collaboration with Shareen Joshi.

20. See Campbell and Shiller (1988). Both series are adjusted for inflation.

The real series of American prices and values are shown in Figure 9 and the simulation results are shown in Figure 10. There is a qualitative correspondence. In both series the price fluctuates around value, and mispricings persist for periods that are sometimes measured in decades. However, at this point no attempt has been made to make forecasts. While I think this is possible, it is beyond the scope of this paper[21]. The point of the above simulation is that a plausible and unsophisticated choice of parameters results in qualitatively reasonable oscillations in the mispricing.

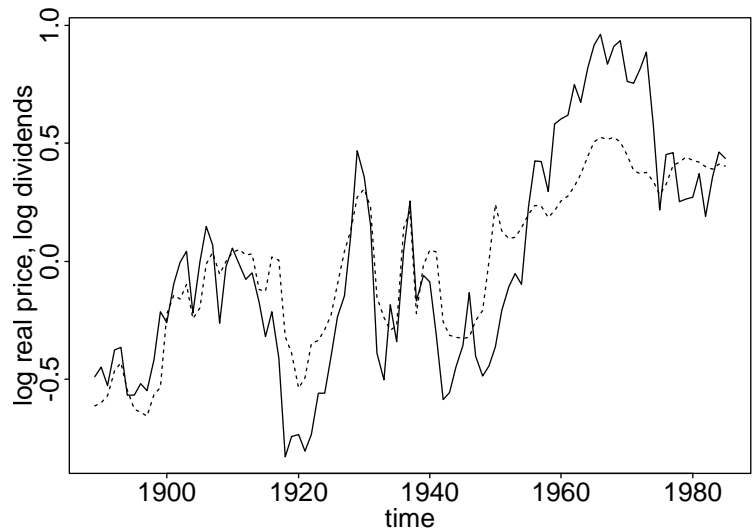

**FIGURE 9. Inflation-adjusted annual prices (solid) and dividends for the S&P index of American stock prices.**

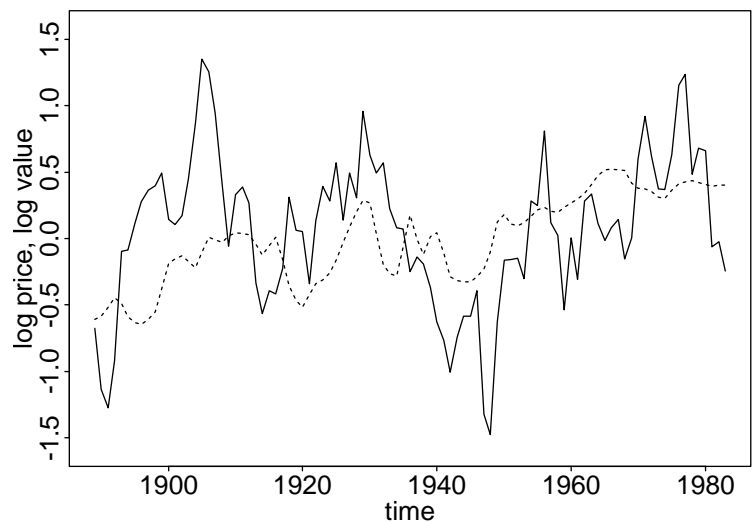

**FIGURE 10. A simulation with value investors and trend followers. The linearly interpolated dividend series from Figure 9 provides the reference value process. Prices are averaged to simulate reduction to annual data. There was some adjustment of parameters, as described in the text, but no attempt was made to match initial conditions. The oscillation of prices around values is qualitatively similar to Figure 9.**

Because of the choice of parameters there is no short term linear structure in this price series. The short term negative autocorrelation of value investors is canceled by the positive short term autocorrelation of trend followers. There is plenty of nonlinear structure, however, as illustrated in Figure 11, which shows the smoothed volume[22] of value investors and trend followers as a function of time. The two groups of traders become active at different times, simply because the conditions that activate their trading are intermittent and unsynchronized. This is true even though the capital of both groups is fixed. Since the trend followers induce positive autocorrelations and the value investors negative autocorrelations, there is predictable nonlinear structure for a trader who understands the underlying dynamics well enough to predict which group will become active. Without knowledge of the underlying generating process, however, it is difficult to find such a forecasting model directly from the timeseries.

Statistical analyses display many of the characteristic properties of real financial timeseries, as illustrated in Figure 11. The log-returns are more long-tailed than those of a normal distribution, i.e. there is a higher density of values at the extremes and in the center with a deficit in between. This also evident in the size of the fourth moment. The excess kurtosis $k = \langle (r_t - \bar{r}_t)^4 \rangle / \sigma_r^4 - 3$ is roughly $k \approx 9$, in contrast to $k = 0$ for a normal distribution. The histogram of volumes is peaked near zero with a heavy positive skew. The volume and volatility both have strong positive autocorrelations. The intensity of the long-tails and correlations vary as the parameters are changed or strategies are altered. However, the basic properties of long tails and autocorrelated volume and volatility are robust as long as trend followers are included.

Clustered volatility has now been seen in many different agent-based models[23]. It seems there are many ways to do produce this behavior. Under the dynamics of equation (8), any strategy with appropriate delays sets up a feedback loop. Large price fluctuations cause large trading volume, which causes large price fluctuations, and so on, generating volatility bursts. Even in the absence of linear structure, the nonlinear structure of a value-trend ecology can cause autocorrelations in volatility. Although oscillations in

---

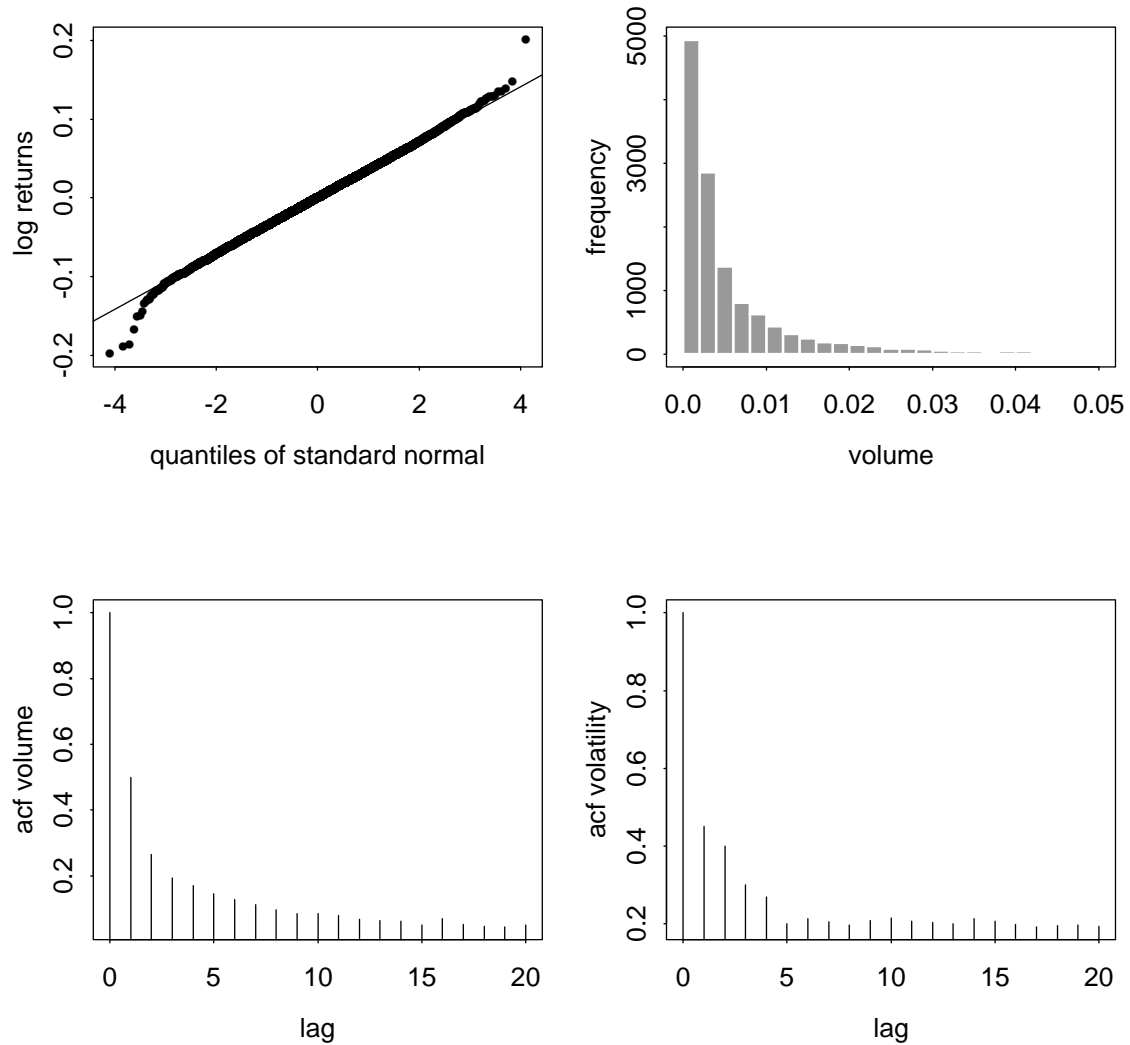

**FIGURE 11. An illustration that an ecology of threshold based value investors and trend followers shows statistical properties that are typical of real financial time series. The upper left panel is a "q-q" plot, giving the ratio of the quantiles of the cumulative probability distribution for the log-returns to those of a normal distribution. If the distribution were normal this would be a straight line, but since it is "fat tailed" the slope is flatter in the middle and steeper at the extremes. The upper right panel shows a histogram of the volume. It is heavily positively skewed. The lower left panel shows the autocorrelation of the volume, and the lower right panel shows the autocorrelation of the volatility. These vary based on parameters, but fat tails and temporal autocorrelation of volume and volatility are typical.**

capital can enhance this, they are not necessary. I believe the vague hand-waving arguments above can be turned into a quantitative theory giving necessary and sufficient conditions for clustered volatility and long tails, but this is beyond the scope of the present paper[24].

The few results presented here fail to do justice to the richness of the trend follower/ value investor dynamics. We have observed many interested effects. For example, the presence of trend followers increases the frequency of the oscillations in mispricing. The mechanism seems to be more or less as follows: If a substantial mispricing develops by chance, value investors become active. Their trading shrinks the mispricing, with a corresponding change in price. This causes trend followers to become active; first the short term trend followers enter, and then successively longer term trend followers enter, sustaining the trend and causing the mispricing to cross through zero. This continues until the mispricing becomes large, but with the opposite sign, and the process repeats itself. As a result the oscillations in the mispricing are faster than they would be without the trend followers. Although in this case the trend followers tend to lose money, in Section 4.3.4 we will give an example in which the reverse is true.

---

24. Stauffer (1999) has suggested that these properties may vanish in the limit $N \rightarrow \infty$. Note that market impact is weighted by capital. If the capital is concentrated in the hands of a few agents, or many agents using only a few strategies, the effective $N$ is small. This is particularly true if capital is distributed according to the Pareto law (see Section 4.2.2).

# 4.   Evolution

In the discussion so far we have treated the capital of each agent or strategy as a fixed parameter. In reality the capital varies as profits are reinvested, strategies change in popularity, and new strategies are discovered. Adjustments in capital alter the financial ecology and change its dynamics, causing the market to evolve. At any point in time there is a finite set of strategies that have positive capital; innovation occurs when new strategies acquire positive capital and enter this set. Market evolution is driven by capital allocation.

Market evolution occurs on a longer timescale than day-to-day price changes. There is feedback between the two timescales: The day-to-day dynamics determine profits, which affect capital allocations, which in turn alter the day-to-day dynamics. As the market evolves under static conditions it becomes more efficient. Strategies exploit profit-making opportunities and accumulate capital, which increases market impact and diminishes returns. The market learns to be more efficient.

In this section we discuss how profits influence market evolution. Use of the log-linear price formation rule simplifies this, since the average profits can be written in terms of aggregate pairwise correlations. This makes it possible to construct a trophic web characterizing the flow of money. Reinvestment of profits leads to a simple model for capital allocation that is a generalization of the standard population dynamics model in biology, with species replaced by strategies, and population replaced by capital.

This can be used to study the progression toward market efficiency. If there are patterns in prices, they should disappear as capital is allocated to strategies that exploit them. This is illustrated by studying a simple example. Myopic blind investment by even a single agent drives profits to zero. The original pattern is eliminated, though a new one can be created in its wake. In contrast, if the agent is smarter and more restrained he will maximize profits. In this case the original pattern is only reduced by half, and thus the market is only partially efficient. If $N$ profit-maximizing agents discover the same pattern and use the same strategy, however, in the limit $N \to \infty$ they fully eliminate the original pattern. In any case the progression toward efficiency is slow: Order of magnitude estimates based on typical rates for capital allocation give timescales to achieve efficiency measured in years to decades.

## 4.1  Flow of money

Financial evolution is influenced by money in much the same way that biological evolution is influenced by food. Profits play a central role. As demonstrated below, the special properties of the log-linear price formation rule lead to simple expressions for profitability, as well as an interesting conservation law.

### 4.1.1  Accounting

The unrealized wealth $w_t^{(i)}$ of agent $i$ is

$$w_t^{(i)} = P_t x_t^{(i)} + u_t^{(i)},$$

where $u_t^{(i)}$ is the money held by agent $i$ at time $t$. The change in money in successive timesteps is

$$u_t^{(i)} - u_{t-1}^{(i)} = P_t(x_t^{(i)} - x_{t-1}^{(i)}) + d_t x_{t-1}^{(i)}$$

The first term on the right is the money needed to buy or sell the asset. The *dividend* $d_t$ allows for the possibility that it might make payments. Putting these two equations together gives the *profit* $g_t^{(i)}$ of the $i^{th}$ agent at time $t$,

$$g_t^{(i)} = (\Delta P_t + d_t) x_{t-1}^{(i)}. \tag{Eq 26}$$

$g_t$ is also called an *unrealized gain*, since it is based on the value of a share marked at the most recent price. This is an optimistic valuation, since conversion to money is risky, and (as we will show later) market impact tends to lower the value.

### 4.1.2  Conservation laws

A financial market is a *closed system* if it does not interact with other financial markets, or with the external economy. This conceptual device makes it possible to discuss conservation laws. There are two obvious conservation laws for financial markets: conservation of shares, and conservation of money. In any given transaction agents exchange shares, and exchange money; while the transaction changes the holdings of individual agents, the totals remain the same. Of course, real markets are *open systems* where new shares can be issued, and there can be net flows of money. For example, an asset can pay dividends, or an agent can import or remove capital. We will see that for there to be any interesting dynamical behavior, the market must be an open system.

The total unrealized wealth is not conserved. To see this, let the market maker's position be $x_t^{(m)}$. Conservation of shares implies that

$$\sum_{i=1}^{N} x_t^{(i)} + x_t^{(m)} = K, \tag{Eq 27}$$

where $K$ is a constant. Multiplying both sides by $\Delta P_{t+1} + d_{t+1}$ and substituting from equation (26), the change in the total wealth at time $t$ is

$$\Delta w_t = \sum_{i=1}^{N} g_t^{(i)} + g_t^{(m)} = (\Delta P_t + d_t)K, \tag{Eq 28}$$

where $g_t^{(m)}$ is the profit of the market maker. Even if we assume $d_t = 0$, generally $\Delta P_t \neq 0$. The total wealth is strictly conserved only if $K = 0$. There are indeed some assets, such as futures and options, where each long position has a corresponding short position, but this is not true in general.

One of the special properties of the log-linear price formation rule is that there is a sense in which *realized wealth* is conserved. Define a *trading cycle* of period $T$ as a situation in which the position $x_{t+T}^{(i)} = x_t^{(i)}$ for all $i$, i.e. all the positions return to some previous value. As a special case, suppose all the directional traders begin and end a cycle with zero positions, realizing all their wealth. Equation (28) makes it clear that the total realized wealth will be conserved providing $d_t = 0$ and $P_{t+T} = P_t$. As a consequence of equation (6), the log-linear rule has the latter property, and so conserves realized wealth. Many temporary equilibrium models also have this property, though this is not true in general[25].

### 4.1.3 Relation between profits and strategies

For financial prices series a typical return $\Delta P_t / P_{t-1}$ is on the order of a percent. Returns are approximately equal to log returns, i.e.

$$r_t = \log P_t - \log P_{t-1} \approx \Delta P_t / P_{t-1}.$$

This is a good approximation when the net order flow $\omega$ is small compared to $\lambda$, and it becomes exact in the limit $\Delta t \to 0$. If we measure the position in units of money ($\tilde{x}_t = P_t x_t$) and the dividend as a fraction of the previous share price ($\tilde{d}_t = d_t / P_{t-1}$), we can rewrite equation (26) as[26]

$$g_t^{(i)} \approx (r_t + \tilde{d}_t) \tilde{x}_{t-1}^{(i)}. \tag{Eq 29}$$

To keep the notation simple going forward we will drop the tildes and hope that the units will always be clear from the context. Substituting for $r_t$ from equation (8) and $\omega_t^{(i)}$ from equation (2),

$$g_t^{(j)} \approx \left( \frac{1}{\lambda} \sum_{i=1}^{N} (x_t^{(i)} - x_{t-1}^{(i)}) + \xi_t + d_t \right) x_{t-1}^{(j)}. \tag{Eq 30}$$

This relation shows how the profits of a given strategy are determined by its relation to other strategies. Under the simplifying assumption that the liquidity is constant when measured in terms of money, as discussed in Section 2.3, by taking time averages and assuming stationarity this can be re-written as

$$\langle g^{(j)} \rangle \approx \frac{1}{\lambda} \sum_{i=1}^{N} G^{(ij)} + \mu^{(j)}. \tag{Eq 31}$$

where

$$G^{(ij)} = \sigma_x^{(i)} \sigma_x^{(j)} (\rho_x^{(ij)}(1) - \rho_x^{(ij)}(0)) \tag{Eq 32}$$

and

$$\mu^{(j)} = \langle (\xi_t + d_t) x_{t-1}^{(j)} \rangle.$$

---

25. Temporary equilibrium models conserve realized wealth if there is a one-to-one correspondence between positions and demands. This seems to be the case in many of agent-based simulations that use temporary equilibrium models, e.g. Levy et al. (1996) and Arthur et al. (1997). In general, however, it is possible that the positions all return to the same value, but the demand of each agent drifts up or down by a constant, leading to a different price. Without such drifts the price is stationary random variable (in contrast to real prices, which are approximately a random walk). In general nonlinear disequilibrium price formation rules generate random walks, even in a fully deterministic setting.

26. Dividends provide an example of how factors other than trading influence prices. When a stock pays a dividend the price typically drops by an amount roughly equal to the dividend. This is generally not driven by selling, i.e. when dividends are paid $\xi_t \approx -d_t$.

$\rho_x^{(ij)}(\tau)$ is the correlation of $x_t^{(i)}$ and $x_{t-\tau}^{(j)}$, and $\sigma_x^{(i)}$ is the standard deviation of $x_t^{(i)}$. The *gain matrix* $G$ describes the profits due to interactions with other strategies, and $\mu^{(j)}$ describes those due to correlations with external fluctuations and the dividend stream. Note that if the market is a closed system then $\mu^{(i)} = 0$ for all $i$.

Equations (31) and (32) provide insight into what makes a strategy profitable. The asymmetric gain matrix $G$ measures the profits of strategy $j$ due to the presence of strategy $i$. These profits increase when strategy $j$ is able to anticipate strategy $i$ (as measured by $\rho_x^{(ij)}(1)$), and decrease if it is similar to strategy $i$ (as measured by $\rho_x^{(ij)}(0)$). The lagged correlation measures whether the strategy is able to *anticipate the majority,* and the contemporaneous correlation measures whether it is in the *minority*. This gives some perspective on recent literature using the El Farol bar problem/minority game, which only has the second term, as a financial model (Arthur 1994, Challet and Zhang 1997). Depending on the sign of $G^{(ij)}$, for any given pair of strategies there are three possible relationships:

- *Competition*: $G^{(ij)} < 0$ and $G^{(ji)} < 0$.

- *Predator-prey:* $i$ preys on $j$ if $G^{(ij)} < 0$ and $G^{(ji)} > 0$.

- *Symbiosis (mutualism):* $G^{(ij)} > 0$ *and* $G^{(ji)} > 0$.

It is surprising that symbiosis is allowed[27].

It is clear from the above derivation that the ability to decompose profits in terms of pairwise correlations is a special property. It is exact for Kyle's linear price formation rule (equation (10)) when positions are in shares, and is a good approximation for the log-linear price formation rule when positions are in units of money. For more general nonlinear price formation rules the profitabilities of individual agents or strategies are more intimately intertwined, and cannot be decomposed pairwise.

### 4.1.4  Market friction

The tendency for uninformed trading to cause losses is called *market friction* by practitioners. For large traders it is the dominant source of transaction costs. Market friction arises because market impact systematically pushes the price in the direction of trading. This is reflected in the fact that the diagonal terms of the gain matrix are negative. This is true because $\rho^{(ii)}(0) = 1$ and $-1 \leq \rho_x^{(ii)}(1) \leq 1$, so from equation (32) it follows that $G^{(ii)} \leq 0$. The case where $G^{(ii)} = 0$ occurs only when $\rho_x^{(ii)}(1) = 1$, which is only possible if the position is constant, implying no trading. Friction increases quadratically with size. This is clear because $G^{(ii)}$ in equation (32) is proportional to $\sigma_x^2$. Friction also depends on trading frequency. For fixed size the friction is proportional to $1 - \rho_x^{(ii)}(1)$; If $\rho_x^{(ii)}(1) \approx 1$, then the position is nearly constant and the turnover is very slow, whereas the highest friction occurs if $\rho_x^{(ii)}(1) = -1$, implying that the position alternates on successive timesteps.

---

27.  Farmer and Joshi (1999) gives an example of symbiotic strategies.

Market friction implies decreasing returns to scale, with a corresponding decrease in profits. To see this, consider a strategy $j$ that is profitable at small scale. For this to be true at least some of the off-diagonal terms of $G$ must be positive. Suppose the scale of this strategy changes according to the transformation $x^{(j)} \to c^{(j)} x^{(j)}$, while all other strategies remain fixed. The standard deviation $\sigma_x^{(j)}$ scales as $c^{(j)}$, while $\rho_x^{(ii)}(\tau)$ remains unchanged. Thus in equations (31) and (32) the positive contributions to the profit from the off-diagonal terms of $G$ grow linearly while the negative contributions from the diagonal terms grow quadratically. The profits reach a maximum and then decline as friction dominates. As a function of capital the profits are an inverted parabola; the return to the strategy, defined as $R^{(i)} = \langle g_t^{(i)} \rangle / c^{(i)}$, decreases linearly. The capital associated with the maximum profit level can be thought of as the *carrying capacity* of the strategy. An example is given in Figure 12.

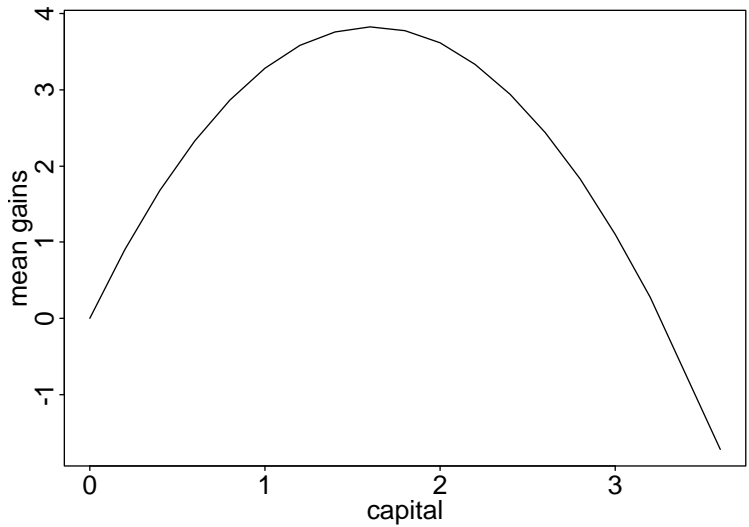

**FIGURE 12. The profit for a trend strategy $\theta = 1$ invading a simple value strategy. (See Section 3.1.1 and Section 3.2). The horizontal axis shows the capital of the trend strategy multiplied by $10^4$. For the value strategy $\alpha = 0.1$. $\sigma_\eta = \sigma_\xi = 0.1$, and $\lambda = 1$.**

This is somewhat simplified, since in general a scale change can alter the interaction of the strategies so that $\rho_x^{(ij)}(\tau)$ is not strictly independent of scale. In numerical experiments I have performed so far this dependence is relatively weak, and the description above seems to be qualitatively correct.

### 4.1.5 Crowding lowers individual profits

A given agent's profits are lowered by other agents using the same strategy. To see this, assume that agent $a$ and $b$ have the same strategy. Then the correlations satisfy

$$\rho = \rho_x^{(aa)}(1) = \rho_x^{(ab)}(1) = \rho_x^{(ba)}(1) = \rho_x^{(bb)}(1),$$

$$\rho_x^{(ab)}(0) = \rho_x^{(ab)}(0) = 1.$$

It is clear from equation (32) that $G^{(ab)}$ and $G^{(ba)}$ are negative providing $\rho < 1$. Thus agents using similar strategies always compete -- each agent's profits are lowered by the presence of the other. For a group of agents using identical strategies, the per-share friction is the same as that of one agent whose capital is equal to the total capital of the group. The per-share profit of each agent is lowered by the presence of the others. This is true even for trend-following strategies.

This is quite different from the situation in which one agent convinces others to follow his lead. If an agent announces his position after it is taken and others follow, this always increases the profit of the first agent. This explicitly *does not* involve agents of the same kind -- the followers are different than the leader. This is true even for trend following strategies; see the remarks in Section 4.3.4.

### 4.1.6 Market maker profits

Under certain conditions market making is guaranteed to be profitable. Assume that the dynamics are stationary, so that time averages are well-defined. Let the market maker's position be $x_t^{(m)}$. The change in the market maker's position is by definition the opposite of the sum of all the directional traders' orders, i.e. $x_t^{(m)} - x_{t-1}^{(m)} = -\sum \omega_t^{(j)}$. Letting $i = m$ in equation (31), pulling the sum inside the average, and substituting in terms of the autocorrelation and standard deviation of the market maker's position (as in the derivation of equation (32)) gives

$$\langle g^{(m)} \rangle = \frac{\sigma_{x^m}^2}{\lambda}(1 - \rho_{x^m}(1)) + \langle (\xi_t + d_t)x_{t-1}^{(m)} \rangle, \qquad \text{(Eq 33)}$$

where $\sigma_{x^m}^2$ and $\rho_{x^m}(1)$ are the variance and first autocorrelation of the market maker's position. Since $-1 \le \rho_{x^m}(1) \le 1$, the first term is positive unless $\rho_{x^m}(1) = 1$. Thus market making is profitable as long as the market maker's position is uncorrelated to the sum of the noise process and the dividend stream. These profits are driven by the aggregate of the directional traders' losses due to market friction. This is clear from equation (28), since absent dividends or drifts in the price, taking averages gives

$$\langle g_t^{(m)} \rangle = - \sum_{i=1}^{N} \langle g_t^{(i)} \rangle . \qquad \text{(Eq 34)}$$

## 4.2 Evolutionary dynamics

### 4.2.1 Capital allocation

Decisions about capital allocation are made by human beings, which makes them difficult to model exactly. However, the human fondness for money and the tendency to rely on strategies that have been successful in the past introduce regularities. These can be described by dynamical equations that are equivalent to a standard model in population biology.

A few of the important factors that influence capital allocations are:

1. *Reinvestment of earnings.* A fraction of profits are added to existing capital.

2. *Attracting capital from investors.* For example, money managers solicit capital from outside investors in return for a share of the profits.

3. *Economic necessity.* Strategies can have utility that is not based on expected profit. Examples are liquidation of assets for consumption, or trading that is motivated by risk reduction.

4. *Capacity limitations.* Because of market friction there is a size where profits reach a maximum, as discussed in Section 4.1.4. Competent traders attempt to attain this maximum by limiting their capital. Successful money managers close their funds when friction becomes too large.

A simple model of reinvestment captures features of the first three factors above.

$$\Delta c_t^{(i)} = c_t^{(i)} - c_{t-1}^{(i)} = a^{(i)} g_{t-1}^{(i)} + \gamma^{(i)} . \qquad \text{(Eq 35)}$$

As before, $c_t^{(i)}$ is the capital for agent $i$ at time $t$, and $g_{t-1}^{(i)}$ is the profit. The profit is lagged by one because it is not available for reinvestment until the following period. $a^{(i)}$ is the reinvestment rate. Because of the ability to attract outside capital, which depends on past performance, it is possible that $a^{(i)} > 1$.

$\gamma^{(i)}$ models the rate at which a strategy attracts capital for reasons unrelated to profitability. This might be because it serves another function, such as risk reduction, or because it appeals to human psychology. Since directional traders tend to lose money as a group, absent any inflows of money their total capital will go to zero, and the dynamics will settle into a trivial fixed point. For the market to sustain itself there must be an outside source of money, either dividends or positive $\gamma^{(i)}$ terms. The market must be an open system to have any interesting behavior.

Equation (35) is just a crude first model of capital allocation. For a discussion for how this is altered by profit maximization see Section 4.4. For other enhancements, such as the effect of opportunity costs, see Farmer and Joshi (1999).

### 4.2.2 Lotka-Volterra dynamics of capital

Combining the results obtained so far gives a set of dynamical equations for the evolution of capital that are analogous to those commonly used to study population dynamics in biology. Suppose we write the positions of each strategy in the form $x_t^{(i)} = c_t^{(i)} \tilde{x}_t^{(i)}$, where $\tilde{y}$ is a scale independent form of the strategy. Substituting equation (30) into equation (35) allows us to write the time evolution of the capital as a dynamical system.

$$\Delta c_{t+1}^{(j)} = \sum_{i=1}^{N} A_t^{(ij)} c_t^{(i)} c_{t-1}^{(j)} - B_t^{(ij)} c_{t-1}^{(i)} c_{t-1}^{(j)} + \mu_t^{(j)} c_{t-1}^{(j)} + \gamma^{(j)} , \qquad \text{(Eq 36)}$$

where

$$A_t^{(ij)} = \frac{a^{(j)}}{\lambda} \tilde{x}_t^{(i)} \tilde{x}_{t-1}^{(j)} \qquad B_t^{(ij)} = \frac{a^{(j)}}{\lambda} \tilde{x}_{t-1}^{(i)} \tilde{x}_{t-1}^{(j)} \qquad \mu_t^{(j)} = (\xi_t + d_t) \tilde{x}_{t-1}^{(j)} .$$

This is a quadratic difference equation with time varying coefficients.

If reinvestment is sufficiently slow then this can be written in a simpler and more familiar form. When $a$ is small the capital changes slowly. The mean profit $g_t^{(i)}$ in a trailing window over $n$ iterations is

$$\frac{1}{n} \sum_{j=1}^{n} g_{t-j}^{(i)} \approx \langle g^{(i)}(\check{c}) \rangle ,$$

where $\langle g^{(i)}(\check{c}) \rangle$ is the time average of the profit with the vector of capitals $\check{c} = (c^{(1)}, ..., c^{N})$ fixed at their trailing mean values

$$c^{(i)} = \frac{1}{n} \sum_{j}^{n} c_{t-j}^{(i)} .$$

$\langle g^{(i)}(\check{c}) \rangle$ can be further decomposed as in equations (31) and (32). By writing each strategy in the scale independent form $x^{(i)} = c^{(i)} \tilde{x}^{(i)}$, equation (35) becomes a differential equation of the form

$$\frac{dc^{(j)}}{dt} = \frac{a^{(j)}}{\lambda} \sum_{i=1}^{N} \tilde{g}^{(ij)} c^{(i)} c^{(j)} + \tilde{\mu}^{(j)} c^{(j)} + \gamma^{(i)} , \qquad \text{(Eq 37)}$$

where

$$\tilde{g}^{(ij)} \;=\; \sigma_x^{(i)}\sigma_x^{(j)}(\rho_x^{(ij)}(1)-\rho_x^{(ij)}(0))$$

and

$$\tilde{\mu}^{(j)} \;=\; \langle(\xi_t+d_t)\tilde{x}_{t-1}^{(j)}\rangle\,.$$

These are the *generalized Lotka-Volterra equations* (e.g. Murray 1990, Hofbauer and Sigmund, 1998), which are the standard model of population dynamics. For financial markets the population is replaced by the capital. Because $\tilde{g}$ is defined in terms of scale-independent quantities it typically varies slowly with $\check{c}$, and for many purposes it is reasonable to treat it as a constant.

The Lotka-Volterra equations were originally introduced to explain oscillations in populations with predator-prey relationships. The dynamics can be unstable or stable, with fixed points, limit cycles, or chaotic attractors. The condition for fixed points is obtained by setting the left hand side of equation (37) to zero, which gives a system of $N$ coupled quadratic equations for $c^{(i)}$ with $N$ unknowns. Such equations can have as many as $2^N$ roots and the possibility of multiple equilibria. Numerical experiments in Farmer and Joshi (1999) show interesting examples where there appear to be chaotic attractors.

We have so far treated the liquidity as a free parameter. As discussed in Section 2.3, the dynamics depend only on the non-dimensional ratios $\alpha^{(i)} = c^{(i)}/\lambda$, so scaling all the capitals by a constant is equivalent to changing $\lambda$. When these equations have a unique attractor, the choice of $\lambda$ becomes irrelevant once initial transients have died out.

Solomon and Levy (1996) and Malcai et al. (1998) have argued that the discrete generalized Lotka-Volterra equations with random linear driving term ($\mu_t^{(j)}$ in equation (36)) give rise to power law distributions. This provides a possible explanation for the Pareto distribution of wealth. However, their usage of the Lotka-Volterra equations in this context was strictly *ad hoc.* The derivation given here lends support to the usage of the Lotka-Volterra equations as a model of relevance for financial economics. In the limit of rapid reinvestment their assumption that $\mu_t$ is random may be reasonable. If so this suggests that the power law concentration of wealth is purely a matter of luck, caused by the self-reinforcing nature of an exponential amplification process with frictional terms.

The model developed here can be connected to evolutionary game theory (e.g. Weibull 1996, Samuelson 1998). The fact that the profits can be understood as the aggregate of the pairwise interactions of the strategies is a useful simplification. Hopefully this will stimulate more realistic applications of evolutionary game theory to problems in finance.

## 4.3  Efficiency, diversity, and learning

### 4.3.1  The evolutionary view of the progression toward market efficiency

In their introductory textbook, Sharpe et al. (1995) define market efficiency[28] as follows:

> *"A market is efficient with respect to a particular set of information if it is impossible to make abnormal profits by using this set of information to formulate buying and selling decisions."*

The basic idea underlying the theory of efficient markets is that the act of exploiting a pattern to make a profit alters the market so that the pattern disappears. The intuition is articulately expressed by Cootner (1964):

> *"If any substantial group of buyers thought that prices were too low, their buying would force up the prices. The reverse would be true for sellers... the only price changes that would occur are those that result from new information. Since there is no reason to expect that information to be non-random in appearance, the period-to-period price changes of a stock should be random movements, statistically independent of one another."*

Samuelson (1965) proved that "properly anticipated prices fluctuate randomly" by assuming that all agents make perfect forecasts of the mean price. Efficiency can also be derived from first principles by assuming the existence of a rational expectations equilibrium (e.g. Lucas 1972). Deviations from these assumptions lead to deviations from perfect efficiency. For example, Grossman and Stiglitz (1980) showed that this occurs for asymmetric information, and there are many examples of inefficiency caused by bounded rationality (e.g. Simon 1956, Sargeant 1993).

Most work on market inefficiency is done in an equilibrium framework. By its very nature, the assumption of equilibrium does not address the question of timescale. At equilibrium the market is in a timeless, asymptotic state. Out of equilibrium, in contrast, time is very much present. This leads to an evolutionary view in which the path toward efficiency is incremental. Patterns are eliminated only as new strategies are discovered and their capital increases. In a changing environment there is never enough time for the market to attain perfect efficiency.

Let us sketch the progression toward market efficiency in an out-of-equilibrium context in more detail. Assume that at some point in time, for example after a structural change, there exist predictable patterns in prices, creating niches that can be exploited to make statistically significant profits. Speculators attempt to discover these patterns. But because the situation is new, the data are noisy and limited, and the modeling process is prone to estimation errors, their strategies are imperfect. Nonetheless, if some strategies are profitable their capital will increase. As their capital grows they begin to cancel pre-existing patterns. As discussed in Section 4.4, the pre-existing patterns are diminished. At the same time, each new strategy may also generate *new* patterns, due to non-optimality or side-effects caused by second-order terms in the price-dynamics. The ecology of strategies becomes more complex, and the market becomes more efficient. New structural changes may alter the balance, creating new inefficiencies, so that this process repeats itself many times.

---

28. For reviews see Fama (1970, 1991). For some remarks about the challenges to this theory, see Farmer and Lo (1999) or Zhang (1999).

The complexity of the interactions in a financial ecology can be thought of in terms of a *trophic web,* that can visualized as a directed graph showing who profits from whom. Each node in the graph represents a given agent (or alternatively, a given strategy). Each pair of nodes is linked by an arrow whose direction is determined by the sign of $G^{(ij)} - G^{(ji)}$. Attention can be focused on the most important interactions by imposing a threshold $G_{min}$, so that links with $\left| G^{(ij)} - G^{(ji)} \right| < G_{min}$ are omitted. Nodes that have large positive values of $\gamma^{(i)}$ or profits from dividends are sources of money. The trophic web describes the flow of money from primary users to successful speculators; its size and depth provide a measure of the complexity of the financial ecology. Note that this is well-defined only under the approximation of a linear or log-linear price formation rule, where profits are characterized by pairwise interactions.

To exploit patterns and make the market more efficient, either the individual strategies must become more complex, or the trophic web describing their interactions must become more complex. The balance that emerges depends on the cost of modeling. If modeling is free and easy, and there is an infinite amount of data available for parameter estimation, then any rational agent should find the unique optimal strategy taking all possible patterns into account. But if modeling is expensive agents will be forced to specialize[29]. No single strategy will be able to find and exploit all patterns, and diversity will increase. This will be reflected in the complexity of the trophic web.

### 4.3.2 Timescale for efficiency

If it were possible to fit perfect models on short data sets and allocate arbitrarily large amounts of capital instantaneously, then the timescale for markets to become efficient would indeed be rapid. But anyone who has ever tried to forecast market prices and attract trading capital knows that this is not the case. Prices are noisy, and model fitting is prone to estimation errors. Consequently, to detect a pattern and be reasonably confident that it is real takes time. Similarly, even with exponential growth rates of reinvestment, allocating capital takes time. Both of these factors lead to timescales for efficiency measured in years to decades.

We will first estimate the time needed to test a good trading strategy. A relevant measure of the quality of a strategy is its return-to-risk ratio. This is also called the *information ratio*. In annual units it is the ratio of the expected annual return to the expected standard deviation of annual returns. A strategy with an information ratio of one is generally considered good, and two is considered excellent. Suppose that price returns are normally distributed. For a strategy whose true information ratio is $S$, the expected statistical significance after trading for $t$ years compared with the null hypothesis of random trading is $S\sqrt{t}$. Thus, to test a strategy with $S = 1$ and be confident in it with two standard deviations of statistical significance requires about four years. Nonstationarities and fat tails in returns lengthen this estimate. This justifies the standard rule of thumb that a five year track record is needed to attract significant investment capital.

---

29. This point has also been stressed recently by Capocci and Zhang (1999).

This is further complicated by the "millionth monkey" problem. If one tests 20 random strategies on four years of data, the odds are good that at least one of them will have positive trading results with two standard deviations of significance. Similarly, with a sufficiently large number of fund managers pursuing diverse strategies, the odds become good that some of them will have good track records over any given four year period. This further lengthens the timescale needed to make intelligent capital allocation decisions, and hence the timescale to achieve market efficiency.

For the market to become efficient capital has to be allocated to good strategies. If efficiency is achieved through pure reinvestment of profit, for a fund with a rate of return of 25% per year, it takes roughly 10 years to increase the funds under management by an order of magnitude. Thus starting from a million dollars and increasing to a billion dollars takes 30 years. Attracting capital from outside sources can shorten this considerably; for example, if the capital doubles every year this is reduced to 10 years.

Thus we see that arguments based on either reinvestment rates or on statistical significance lead to timescales measured in years to decades. The progression toward efficiency is necessarily slow.

### 4.3.3 Increasingly rational expectations and evolution toward higher complexity

To make a connection to the rational expectations framework, imagine there existed a *super-agent* who knew the strategies and capital levels of all other agents, fully understood the price formation process, and had access to all information sources. If this agent were the only one with these powers, he could iterate equation (8), make an optimal prediction of future prices, and use this to make profits[30]. Since the strategy of the super-agent involves simulating all other strategies in the market, its algorithmic complexity[31] is necessarily greater than or equal to that of all other strategies combined.

Now suppose there were a second super-agent, unknown to the first, with knowledge of all strategies and capital levels, including those of the first super-agent. As before, super-agent II could simulate the market, including the behavior of super-agent I, and make profits. One can imagine a sequence of such agents, each with successively more knowledge. The strategies of these agents would have ever-increasing algorithmic complexity.

If *all* the agents are super-agents, with perfect knowledge of each other, I conjecture that this leads to a rational expectations equilibrium. Game theoretic treatments of markets, e.g. Dubey et al. (1987), support this view. The scenario envisioned above makes it possible to think of a rational expectations equilibrium as a limit in which each agent acquires successively more knowledge of the other agents. I suspect that the algorithmic complexity needed to actually reach this limit is infinite, even though it may be quite small at the limit itself. Thus while the strategies needed to implement rational expectations equilibrium may be very simple, the strategies needed to get there are not.

---

30. Because the spread is assumed to be zero, for any level of predictability there is a capital level small enough to guarantee profitability.

31. The algorithmic complexity is the length of the shortest program capable of implementing a given function.

Of course, it is unrealistic to assume that an agent could know the strategies of all other agents in detail. Good traders go to great pains to keep their strategies secret, so that any knowledge of the strategies of others is at best vague. There can be no super-agents in real markets. Nonetheless, the ability of agents to infer the strategies of others may cause a progression toward greater complexity, including red-queen effects, so that strategies have to become more and more complicated just to stay even. (See e.g. Browne (1995)).

### 4.3.4 Technical trading as a simple way to exploit inefficiencies

In a perfectly efficient market, the use of technical trading strategies is a highly irrational, risky way to lose money. Nonetheless, market surveys make it clear that technical strategies, of which trend strategies are just one example, are widely used (Keim and Madhaven, 1995; Menkoff 1998). There is some evidence in the literature that there have been periods where technical strategies made profits (Brock et al. 1992, LeBaron 1998). This section uses trend strategies to illustrate how technical trading strategies can provide a cheap, simple way to exploit inefficiencies created by other strategies[32].

The calculations are made simpler by using a standard technique from population genetics. If a new strategy is introduced into a pre-existing market with only a small amount of capital, if it makes a profit, then according to equation (37) its capital will increase, and it will *invade* the population. The calculation of profitability is simplified by the assumption that the capital of the invading strategy is small enough so that it has a negligible effect on pre-existing price dynamics.

When do trend strategies make profits? From equation (29), providing the capital $c^{(i)}$ of the trend following strategy is negligible, the average profits are

$$\langle g_{t+1}^{(i)} \rangle \approx \langle r_{t+1} x_t^{(i)} \rangle = c \langle r_{t+1}(p_{t-1} - p_{t-\theta-1}) \rangle = c \sum_{i=1}^{\theta} \langle r_{t+1} r_{t-i} \rangle.$$

As long as $r_t$ is stationary with $\langle r_t \rangle = 0$, the mean return $\langle R^{(i)} \rangle = \langle g_t^{(i)} \rangle / c^{(i)}$ to the trend following strategy is

$$\langle R^{(i)} \rangle \approx \sigma_r^2 \sum_{i=1}^{\theta} \rho_r(i+1), \tag{Eq 38}$$

where $\sigma_r^2$ and $\rho_r$ are the variance and autocorrelation of the log-returns in the absence of the trend strategy. Thus the profits of this simple trend strategy with timescale $\theta$ are proportional to the sum of the autocorrelations of the price returns from $\tau = 2$ to $\theta$.

---

32. The ability of technical strategies to generate profits has also been emphasized by Brock and Hommes (1997, 1998, 1999).

There are many "fundamental" causes for positive autocorrelations in prices, creating opportunities for trend strategies. Some of them include gradual acquisition of large positions to minimize market impact, chain reactions of buying or selling due to placement of nearby stop loss orders or option strike prices, slow diffusion of information, and gradual unloading of large market maker positions accumulated due to random variations in order flow.

Another cause of positive correlations are other strategies that induce them in the price. Several examples were given in Section 3. For example, a simple value investing strategy induces autocorrelations of the form $\rho_r(\tau) = (-\alpha)^\tau$. The second autocorrelation $\rho_r(2) = \alpha^2$ is positive. In a market dominated by simple value investors, equation (38) makes it clear that a trend strategy with $\theta = 1$ will make profits, as shown in Figure 12.

Trend strategies also induce positive autocorrelations. It is a widely held but erroneous view that, when using a trend strategy, it is to one's advantage if others use the same trend strategy. The discussion given in Section 4.1.5, makes it clear that this is false. To check this, note that according to equation (38) and equation (25) the returns when a trend strategy with $\theta = 1$ tries to invade itself are proportional to $\rho(2) = -\alpha/(1 + \alpha) < 0$. However, trend strategies with short timescales can invade those with longer timescales. As computed in Section Section 3.2, for large $\theta$ a trend strategy induces positive autocorrelations for $1 \leq \tau \leq \theta/2$, and the running sum is positive for $\tau < \theta$. Thus a trend strategy with lag $\theta' < \theta - 1$ can invade a trend strategy with timelag $\theta$; the profits are maximized when $\theta' = \theta/2$. In contrast, a trend strategy with $\theta' > \theta$ takes losses, and cannot invade.

It is instructive to compare to the work of DeLong et al. (1990). They show that when a rational investor exploits trend followers, he may amplify trends in the price and slow the process of price adjusting to value. There are several differences between their treatment and mine. Since they use the equilibrium framework, to keep their model tractable it is formulated in terms of only two trading periods. As a result they did not observe the longer-term aspects of the price dynamics, such as oscillatory behavior. The other major difference is in the assumptions about the knowledge and capabilities of agents that are integral to the rational expectations framework. In their model, trend followers are assumed to be irrational noise traders, who lose money on average. In contrast, their rational investors have perfect knowledge of the correct price function, which implicitly involves the strategies and capitals of all the other agents in the market. The underlying realism of such an assumption is doubtful. Technical strategies, in contrast, involve information that is available to everyone. While their work supports the notion that rational investors may appear to be trend followers, this analysis takes this a step farther, by showing the circumstances in which trend strategies are profitable.

## 4.4 Pattern evolution

If markets are efficient, then when a pattern is discovered it should disappear as the capital of profitable trading strategies grows. Understanding the necessary conditions for this in the general case is difficult, since there is a cascade of effects: The new trading alters prices, which alter trading, which alters prices, etc. This is complicated and is beyond the scope of this paper. But it is relatively easy to solve this problem for the special case of an isolated pattern, as done in this section.

To understand why blind investment is not an adequate model, consider a trader with a strategy that is profitable at small size. The capital of this trader will grow until the average profit reaches a maximum. When this occurs, there are two possibilities. If the trader fails to understand market impact, he will continue to blindly reinvest his profits, lowering the returns until they go to zero (see Figure 12). The original pattern is fully eliminated, and the market becomes efficient. But if he understands market impact, he will maximize profits by limiting his capital, and equation (37) ceases to apply. In this case, as we will show, the pattern is reduced by half. The market does not become efficient.

### 4.4.1 The effect of market impact on an isolated pattern in prices

To keep things simple, and to make the results easy to visualize, we analyze the special case of a temporally isolated pattern, of the form

$$\Pi = (\ldots, 0, \pi_{t+1}, 0, \ldots),$$

where $\pi_{t+1}$ is the mean return at time $t+1$. The mean can be defined in terms of an ensemble average of the form

$$\pi_{t+1} = \int r_{t+1} P(r_{t+1}) dr_{t+1},$$

where $P(r_{t+1})$ is the probability density of the return $r_{t+1}$. We will assume that to exploit this pattern a new trader takes a new position $c$ at time $t$ and exits it at time $t+1$. (Exiting immediately is natural since it minimizes risk.) The assumption that the pattern is isolated simplifies the calculation, since it means that the prices for times $t' < t$ are unaltered by the new trading. Defining the pattern in terms of unconditional averages is somewhat misleading, since it does not address the issue of what is knowable. A better approach would be to define patterns in terms of the conditional probability based on a given information set. This complicates the calculations considerably. Simply computing the unconditional mean is simpler, and gives an upper bound on predictability.

The new pattern, which includes the market impact of the new trading, is of the form $\tilde{\Pi} = (\ldots, 0, \tilde{\pi}_t, \tilde{\pi}_{t+1}, \tilde{\pi}_{t+2}, \ldots)$. $\tilde{\pi}_t$ is nonzero because of the new trading, and $\tilde{\pi}_{t+2}, \tilde{\pi}_{t+3}, \ldots$ are nonzero because of the cascade of price alterations that the new trading generates. The new pattern can be computed by assuming that $c$ is small, and that the trading strategies that generated the original pattern are smooth functions of prices that can be expanded in a Taylor's series. The calculation is done in the Appendix. As illustrated in Figure 13, the result is

$$\tilde{\pi}_t = \frac{c}{\lambda}$$

$$\tilde{\pi}_{t+1} \approx \pi_{t+1} - (1 - \kappa_t^0)\frac{c}{\lambda} \quad , \qquad \text{(Eq 39)}$$

$$\tilde{\pi}_{t+2} \approx (\kappa_{t+1}^0 \kappa_t^0 + \kappa_{t+1}^1)\frac{c}{\lambda}$$

where

$$\kappa_t^j = \frac{1}{\lambda} \langle \sum_{i=1}^{N} \frac{\partial \omega^{(i)}(p_t, p_{t-1}, \dots)}{\partial p_{t-j}} \rangle . \qquad \text{(Eq 40)}$$

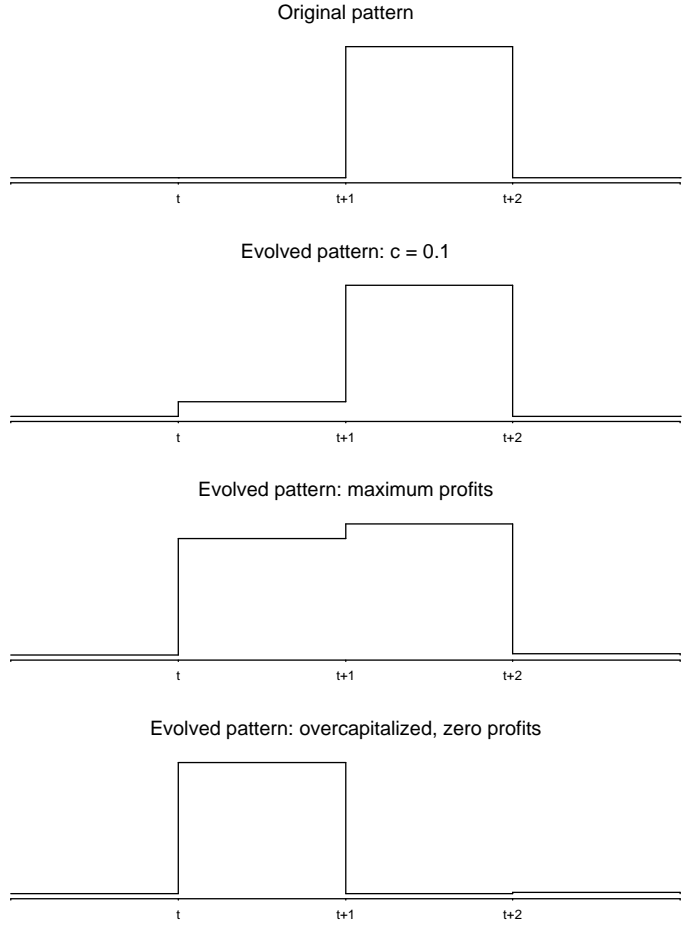

**FIGURE 13. The evolution of an isolated pattern as it is exploited with increasing capital. The price sensitivities are** $\kappa_t^0 = \kappa_{t+1}^0 = -0.1$ **and** $\kappa_{t+1}^1 = 0$. **As the capital is increased to** $c = 0.1$, **the pattern is diminished at time** $t+1$ **and enhanced at time** $t$. **As** $c$ **is increased this trend continues. The profit is maximized at** $c \approx 0.45$, **and the pattern is spread between** $t$ **and** $t+1$. **If the strategy is over-capitalized to the point that profit goes to zero, the original pattern is entirely shifted to the previous timestep.**

Similar expressions are possible for $\tilde{\pi}_{t+3}$, $\pi_{t+4}$, etc., but as long as $\left|\kappa_t^j\right| \ll 1$ the disturbance diminishes as $t$ increases.

$\kappa_t^j$ is the *price sensitivity*. It describes the sensitivity of the price at time $t+1$ to changes in the price at time $t-j$. Since this depends on the sum over all the strategies, we could also consider the sensitivity of each strategy separately. On average value investing strategies have negative price sensitivity, and trend following strategies have positive price sensitivity. The new pattern at time $t+1$ is smaller than the old pattern providing the total price sensitivity $\kappa_t^0 < 1$. This is precisely the condition for linear stability.

Figure 13 shows how the new pattern evolves as it is exploited more and more by increasing $c$. It initially becomes smoother, as reflected by the fact that $\left|\tilde{\pi}_t\right| \geq 0$ and $\left|\tilde{\pi}_{t+1}\right| < \left|\pi_{t+1}\right|$. It eventually reaches the point where the pattern at $t$ is stronger than the pattern at $t+1$, until the original pattern is entirely gone and the new pattern is fully concentrated at time $t$.

There are two special values of $c$ to consider:

- The value of $c$ where the profit is maximized.

- The value of $c$ where the profit is zero.

The first case is the endpoint of market evolution if the new trader understands market friction and limits capital to maximize profits. The second case is the expected endpoint if profits are blindly re-invested until they go to zero. We can find these special values of $c$ by computing the mean profit

$$\bar{g}(c) \approx \tilde{\pi}_{t+1} c \approx \left(\pi_{t+1} - (1 - \kappa_t^0)\frac{c}{\lambda}\right) c. \qquad \text{(Eq 41)}$$

Assuming $\kappa_t^0 < 1$, as a function of $c$ this is an inverted parabola whose maximum is determined by the condition $\partial \bar{g} / \partial c = 0$. This occurs when

$$c = \frac{\lambda \pi_{t+1}}{2(1 - \kappa_t^0)}.$$

The mean profit at the maximum is approximately

$$\bar{g}_{max} = \frac{\lambda \pi_{t+1}^2}{4(1 - \kappa_t^0)},$$

and the evolved pattern is approximately

$$\tilde{\pi}_t = \frac{\pi_{t+1}}{2(1 - \kappa_t^0)}$$

$$\tilde{\pi}_{t+1} = \frac{\pi_{t+1}}{2}$$

$$\tilde{\pi}_{t+2} = \frac{(\kappa_{t+1}^0 \kappa_t^0 + \kappa_{t+1}^1)\pi_{t+1}}{2(1 - \kappa_t^0)}$$

If the trader maximizes profits the pattern at time $t + 1$ is exactly half as large as it was before. In contrast, if the trader simply blindly reinvests, the profit goes to zero. The new pattern in this case is

$$\tilde{\pi}_t = \frac{\pi_{t+1}}{(1 - \kappa_t^0)}$$

$$\tilde{\pi}_{t+1} = 0$$

$$\tilde{\pi}_{t+2} = \frac{(\kappa_{t+1}^0 \kappa_t^0 + \kappa_{t+1}^1)\pi_{t+1}}{(1 - \kappa_t^0)}$$

This is illustrated in Figure 13. We see that as the capital is increased the pattern evolves earlier in time. If the trader blindly reinvests the pattern is entirely shifted to the previous timestep.

Figure 14 shows how the price sensitivity affects the pattern in the case where $c$ is chosen to maximize profits. The new pattern at time $t + 1$ is half as big as it was before, independent of the price sensitivities. The new patterns at time $t$ and time $t + 2$, however, depend on the price sensitivity. If the price sensitivity is zero, $\tilde{\pi}_t$ is half the size of the original pattern, but it is greater than half if the price sensitivity is positive, and less than half if it is negative. Patterns with positive price sensitivity are more difficult to remove than those with negative price sensitivity. Thus patterns created by trend following are more resistant to market efficiency than those created by value investing.

The most important result of this section is that if a single agent discovers a pattern and maximizes profits, the original pattern is reduced by roughly half. Thus a single agent maximizing profits does not make the market efficient all by himself. This is in contrast to blind reinvestment, which drives profits to zero. In my experience as a practitioner, understanding market impact well enough to accurately limit capital is a difficult statistical estimation problem that even the best traders have a difficult time solving. The statistical errors in estimating expected profitability are so large that it is easy to be off by a factor of two. The assumption that financial agents behave as profit maximizers, as asserted by Friedman (1953), is highly questionable[33]. It is not obvious whether profit maximization or blind reinvestment provide the better model of reality.

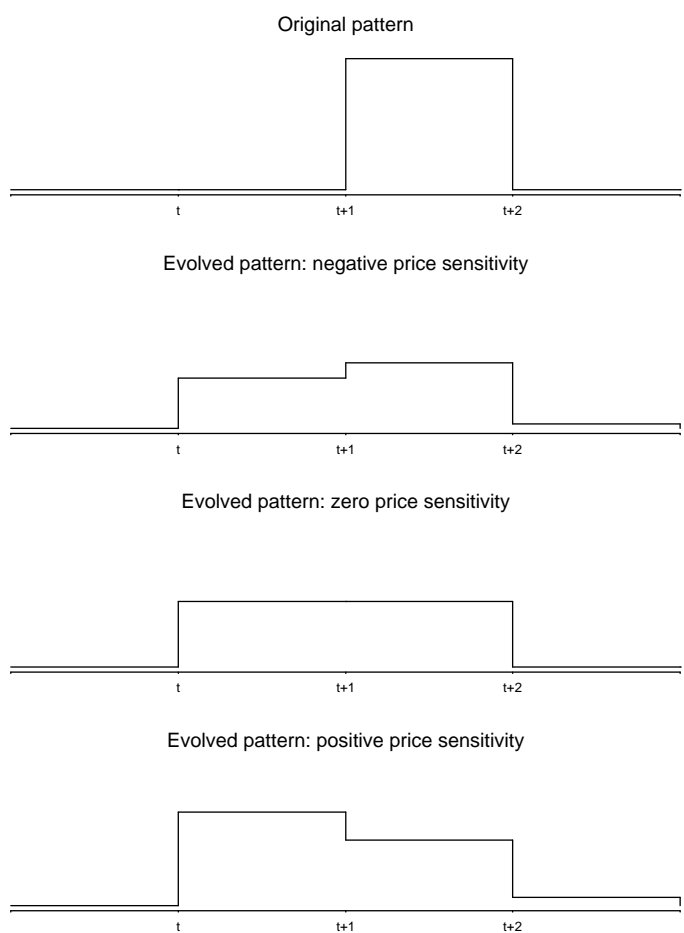

**FIGURE 14. The effect of the price sensitivity on pattern evolution. Assume the trader adjusts his capital to maximize profit. The pattern at time $t + 1$ evolves to half its original size, independent of the price sensitivity. At time $t$ it is less than, equal to, or greater than half the size of the original, depending on whether the price sensitivity is negative, zero, or positive.**

### 4.4.2 Many traders in the same niche

Up until now there has been little discussion of what happens when agents share strategies. In terms of market impact there is no difference between one trader using a strategy, and ten traders using the same strategy, each with a tenth of the capital. However, the coordination problem for agents to maximize profits, and the resulting endpoint of market evolution, are quite different. A single trader maximizing profits will diminish the pattern he is exploiting by half. In contrast, if there are $N$ traders using the same strategy, each independently maximizing individual profits, the pattern goes to zero in the limit $N \to \infty$.

---

33. See Blume and Easley (1992, 1998) for related previous work. They make the important distinction between behavior that is *sufficing* (good enough to survive) vs. optimal. In the context of an intertemporal equilibrium model with capital markets included they show that market selection can lead to complicated dynamics, and does not always lead to profit maximizing behavior.

Thus to achieve market efficiency under profit maximization it is necessary that many traders occupy the same niche.

This is illustrated for the case of an isolated pattern discussed in the previous section. Instead of a single new trader, assume there are $N$ new traders, each of whom make trades $(c_i, -c_i)$ at times $(t, t+1)$. The mean profit for trader $i$ is similar to equation (41), with $c$ replaced by $\sum c_i$ for the return, but $c_i$ for the position.

$$\bar{g}_i \approx \left( \pi_{t+1} - \frac{(1 - \kappa_t^0)}{\lambda} \sum_{j=1}^{N} c_j \right) c_i.$$

The maximum profit for trader $i$ occurs when

$$\frac{\partial \bar{g}_i}{\partial c_i} = 0 = \pi_{t+1} - \frac{(1 - \kappa_t^0)}{\lambda} \left( \sum_{j=1}^{N} c_j + c_i \right).$$

Applying this to each trader $i = 1, \dots, N$, the corresponding capital where the profits are maximized is

$$c_i = \frac{\lambda \pi_{t+1}}{(N+1)(1 - \kappa_t^0)}.$$

The new pattern at time $t+1$ is

$$\tilde{\pi}_{t+1} = \frac{\pi_{t+1}}{N},$$

and the resulting maximum profit for each trader is

$$g = \frac{\lambda \pi_{t+1}^2}{N(N+1)(1 - \kappa_t^0)}.$$

Thus as $N$ grows larger the original pattern rapidly disappears, and the total profit associated with the strategy goes to zero. This is in marked contrast to what would occur if the agents were to cooperate, and limit their capital so that

$$c_i = \frac{\lambda \pi_{t+1}}{2N(1 - \kappa_t^0)}.$$

In this case their individual profits are

$$g = \frac{\lambda \pi_{t+1}^2}{4N(1 - \kappa_t^0)},$$

i.e. they evenly divide the profits that would have been made with $N = 1$. They are clearly much better off if they cooperate and collectively limit their capital.

Thus we see that trading resembles the prisoner's dilemma and other games in which the competitive equilibrium results in lower utility than the cooperative equilibrium. When agents control their capital to maximize profits, the number of agents becomes important. Assuming each agent maximizes profits independently, the market becomes efficient in the limit as the number of agents goes to infinity.

## 4.5  Analogy to biology

The analogy between a biological ecology of interacting species and a financial ecology of interacting strategies is summarized in Table 3.

| Biological ecology | Financial ecology |
| --- | --- |
| species | trading strategy |
| individual organism | trader |
| genotype | functional representation of strategy |
| phenotype | actions of strategy (buying, selling) |
| population | capital |
| external environment | price and other informational inputs |
| free energy | money |
| niche | a possible flow of money |
| individual organisms | traders |
| plants | strategies that have utility other than profits |
| animals | speculators |
| selection | capital allocation |
| mutation and recombination | creation of new strategies |

**TABLE 3. An analogy between biological and financial ecologies.**

The idea that there are analogies between biology and economics has a long history (for reviews see Nelson and Winter, 1982, England 1994, or Ruth 1996). The analogy developed here is similar to that implicit in treatments of economics based on evolutionary game theory (Samuelson 1998, Weibull 1996, Friedman 1999). This analogy is by no means exact, but expressing it mathematically makes the similarities and differences clearer. For example, I have identified trading strategies with species, and the capital invested by a trading strategy with population. This is given substance by the generalized Lotka-Volterra dynamics developed in Section 4.2.2.

Some of the elements of the analogy as stated in Table 3 are not as solid, and are included to be thought-provoking. For example, I have identified free energy with money. Both are sought after by individual agents, and play an important role in modulating selection. Both make it possible to "do things"; in biology, free energy enables purposeful action, and in finance money makes investments possible. In this vein, in finance a niche is an opportunity for a strategy to exist, either because of a potential profit making opportunity, or because of some other utility that enables an inflow of money from outside the market (positive $\mu$ or $\gamma$ in equation (37)). Strategies maintained by inflows of money are like plants, which enable the fixation of free energy, and speculators are like animals.

Table 3 stresses the similarities between biological and financial ecologies, but there are also many differences:

- There is no obvious analogy to prices. In most modern financial markets the prices of transactions are broadcast almost instantaneously. This provides a strong constraint on the interaction of financial agents that has no clear analogy in biology.

- The innovation process in financial markets is quite different.

  - There is no inter-species breeding restriction for financial strategies. As a consequence of this, one would expect the taxa of financial strategies to vary continuously. However, there are several mechanisms that may cause clustering, such as cultural broadcasting and specialization (driven by costs of knowledge and information).

  - Financial innovation is Lamarckian, in the sense that necessity is the mother of invention. The pace of innovation for directed human problem solving is presumably much faster than that of random variation. People can reason *de novo,* at least in principle. The consequences of this are discussed more fully in the conclusions.

# 5. Conclusions

## 5.1 Summary of results

I have tried to show how a simple non-equilibrium price formation rule and a simple capital allocation rule can be used to study market evolution. While neither of these are fully realistic, they may be the simplest choices that give reasonable results. They can be written in the form

$$p_{t+1} = p_t + \frac{1}{\lambda} \sum_{i=1}^{N} (c_{t+1}^{(i)} \tilde{x}_{t+1}^{(i)} - c_t^{(i)} \tilde{x}_t^{(i)}) + \xi_{t+1}$$

$$c_{t+1}^{(i)} = c_t^{(i)} + a^{(i)} g_t^{(i)} + \gamma^{(i)}$$

$$g_t^{(i)} = (P_t - P_{t-1} + d_t) c_{t-1} \tilde{x}_{t-1}^{(i)}$$

(Eq 42)

The first equation is the log-linear price formation rule with additive noise $\xi_t$, written with the function $x_t^{(i)}$ giving the position of agent $i$ separated into a scale independent term $\tilde{x}_t^{(i)}$ and the capital $c_t^{(i)} > 0$. $p_t$ is the logarithm of the price and $\lambda$ is the liquidity. The second equation is the blind reinvestment capital allocation rule. The profits $g_t^{(i)}$ are reinvested at rate $a^{(i)}$. For nontrivial asymptotic price dynamics, there must be an inflow of capital, either because the asset pays positive dividends $d_t$, or because $\gamma^{(i)} > 0$ for at least some values of $i$. The third equation is the standard accounting rule for profits, written in terms of the price $P_t$.

It is necessary to specify the functions $\tilde{x}_{t+1}^{(i)} = \tilde{x}(p_t, p_{t-1}, ..., I_t^{(i)})$, which give the decision rule or strategy for each agent. In Section 3 I have investigated the price dynamics of several value and trend strategies with fixed capital ($a^{(i)} = \gamma^{(i)} = 0$). For simple value strategies that take a positive position when an asset is underpriced, and a negative position when it is overpriced, prices do not track values. For some more complicated value strategies with state dependence, prices weakly track values when parameters are in the appropriate range. When there are a diversity of different views about value, the dynamics of linear value strategies can be represented in terms of those of a single representative agent, but for more nonlinear strategies this is not possible, and numerical experiments show an amplification of noise that can be interpreted as excess volatility.

The difference of positions at different times in equation (42) gives rise to second order oscillatory terms that complicate the price dynamics. On a short timescale value strategies induce negative autocorrelations, and trend strategies induce positive autocorrelations. On longer timescales trend strategies induce strong negative autocorrelations, corresponding to oscillations in the price. A combination of value investors and trend followers can result in price dynamics with weak linear structure, but interesting nonlinear structure. This gives rise to qualitatively reasonable boom-bust cycles, long tailed price fluctuations, and clustered volatility.

The log-linear price formation rule has special properties that simplify the flow of money. Equation (42) can be rewritten to express the dynamics of capital in terms of the generalized Lotka-Volterra equations ((36) or (37)). The profits of a given strategy can be decomposed into the sum of its pairwise correlations with other strategies (equation (32) or (37)). Each correlation term an be further decomposed into two parts, one that measures whether a strategy anticipates the majority, and the second whether it is in the minority. Profits are limited by market friction, which is caused by self-impact on the prices. If the capital $c^{(i)}$ of a given strategy is varied while that of all other strategies is held fixed, a strategy either loses for all $c^{(i)}$, or makes profits when $c^{(i)}$ is small, that reach a maximum at some $c^{(i)}_{max}$ and then decline (see Figure 12). Under certain assumptions, the market maker always makes a profit. Another special property of the log-linear rule is conservation of realized wealth. This implies that the profits of the market maker come from market friction. The derivation of the Lotka-Volterra equations makes the analogy between financial markets and biological ecologies precise, and the pairwise decomposition of profits makes it clear how a market can be regarded as an ecology of interacting agents.

The progression toward market efficiency caused by the evolution of capital can be studied within this framework. If there are patterns in prices, the market becomes efficient as new strategies find them and cause them to disappear. Several different estimates lead to a timescale for market efficiency measured in years to decades. For the special case of an isolated pattern I have computed several properties of the progression toward greater efficiency, that illustrate several interesting points. Patterns created by trend following are more difficult to eliminate than patterns caused by value investing. If agents maximize profits, a single agent will only reduce a pre-existing pattern by roughly half, and the market becomes efficient only in the limit that $N$ agents exploit the same pattern.

## 5.2  Closing remarks

In the spirit of Dawkin's (1976) concept of memes, and E.O. Wilson's (1998) program of *consilience,* this paper develops an analogy between financial and biological ecologies. This analogy depends in part on the assumption that financial strategies are *evolving automata. S*trategies can be arbitrarily complex, but the key is that (like a genome) they can be regarded as algorithms that evolve with experience. This is in contrast to the prevailing view in neo-classical economics that the essence of a financial agent is the ability to reason and solve problems *de novo.* Of course, the behavior of real financial agents involves a mixture of both experience and fresh reasoning. My view is strongly influenced by my own experience as a practitioner: The investment strategy used by Prediction Company is based completely on evolving automata. Decisions are made entirely by computers without human intervention, using programs developed through an historical trial and error process. Of course many investors do use *de novo* reasoning, but my guess is that in the vast majority of cases, experience and culture dominate over freshly generated logic. There is a nonetheless a spectrum of possibilities between these two poles: By properly articulating both extremes, and exploring the middle, perhaps we can arrive at a view that correctly characterizes real people in real markets.

## Acknowledgments

Shareen Joshi's help with simulations was invaluable. Sections 3.1 - 3.3 contain a brief summary of joint work that should be published elsewhere. I benefited greatly from discussions and critical comments on earlier drafts of this manuscript from Chris Carroll, Michel Dacorogna, William Finnoff, Dan Friedman, Alan Grafen, John Geanakoplos, Chris Langton, Blake Lebaron, Seth Lloyd, Andrew Lo, Michael de la Maza, Ayla Ogus, Didier Sornette, Spyros Skouras, George Soros, Stuart Trugman, and David Weinberger. I would also like to thank Norman Packard and others at Prediction Company for creating the context that motivated this work, and permitting the leave of absence during which version 3 was finally completed. Thanks are also due to Robert Shiller for making the data used in Section 3.3 available on his web page.

## Appendix

### Derivation of results in Section 4.4.1

This derives the main result in Section 4.4.1. We want to compare an old pattern to a new one that includes the additional trades $(c, -c)$ at times $(t, t+1)$. The new pattern is of the form

$$\tilde{\Pi} = (\ldots, 0, \tilde{\pi}_t, \tilde{\pi}_{t+1}, \tilde{\pi}_{t+2}, \ldots) \quad .$$

Because of the assumption that the old pattern is isolated, the new pattern is unchanged for times $t' < t$. However, for $t' \geq t$ the new trades affect any pre-existing trades, and the impact can propagate arbitrarily far into the future. The impact depends on the aggregate of the trading strategies that created the original pattern. The new pattern is

$$\tilde{\pi}_{t-1} = \pi_{t-1} = 0$$

$$\tilde{\pi}_t = \frac{c}{\lambda}$$

$$\tilde{\pi}_{t+1} = \frac{1}{\lambda}(\langle \omega(\tilde{p}_t, \ldots) \rangle - c) \quad .$$

$$\tilde{\pi}_{t+2} = \frac{1}{\lambda}(\langle \omega(\tilde{p}_{t+1}, \tilde{p}_t, \ldots) \rangle)$$

$$\tilde{\pi}_{t+3} = \ldots$$

That is, for time $t$ the new pattern is entirely caused by the new trading, which produces return $c/\lambda$, and at time $t+1$ it is the sum of that caused by the original trading strategies acting on the altered prices, and that caused by the exiting trade $-c$. Similarly, the new log-prices are

$$\tilde{p}_{t-1} = p_{t-1}$$

$$\tilde{p}_t = p_t + \frac{c}{\lambda}$$

$$\tilde{p}_{t+1} = p_{t+1} + \frac{1}{\lambda}(\omega(\tilde{p}_t, \ldots) - \omega(p_t, \ldots))$$

If $\omega$ is a smooth function whose derivatives exist then providing $c$ is small enough we can approximate it using Taylor's theorem.

$$\tilde{\pi}_{t+1} = \frac{1}{\lambda}\Big( \langle \omega(p_t, \ldots) \rangle + \langle \frac{\partial \omega}{\partial p_t} \delta p_t \rangle - c \Big)$$

$$\tilde{\pi}_{t+2} = \frac{1}{\lambda}\Big( \langle \omega(p_{t+1}, p_t, \ldots) \rangle + \langle \frac{\partial \omega}{\partial p_{t+1}} \delta p_{t+1} \rangle + \langle \frac{\partial \omega}{\partial p_t} \delta p_t \rangle \Big)$$

(Eq 43)

where derivatives are evaluated at the original prices $(p_{t+1}, p_t, \ldots)$ and $\delta p_t = \tilde{p}_t - p_t$. By assumption

$$\pi_{t+1} = \langle \omega(p_t, \ldots) \rangle / \lambda$$

$$\pi_{t+2} = \langle \omega(p_{t+1}, p_t, \ldots) \rangle / \lambda = 0$$

Furthermore, $\delta p_t = c/\lambda$. Using Taylor's theorem again,

$$\delta p_{t+1} = \frac{1}{\lambda}(\omega(\tilde{p}_t, \ldots) - \omega(p_t, \ldots)) \approx \frac{\partial \omega}{\partial p_t} \delta p_t.$$

We can define the price sensitivity as in equation (40), and make the further approximation that

$$\langle \frac{\partial \omega}{\partial p_{t+1}} \frac{\partial \omega}{\partial p_t} \rangle \approx \kappa_{t+1}^0 \kappa_t^0,$$

Collecting these relations together and substituting into equation (43) gives equation (37).